\documentclass[twocolumn,trackchanges]{aastex63}
\usepackage{amsmath}

\usepackage{CJKutf8}

\newcommand{\cntext}[1]{\begin{CJK}{UTF8}{gbsn}#1\end{CJK}\kern-0.8ex}
\newcommand{\RN}[1]{\textup{\uppercase\expandafter{\romannumeral#1}}}
\usepackage{color}

\turnoffedit

\received{TBD}
\revised{TBD}
\accepted{TBD}
\submitjournal{ApJ}

\shortauthors{Yu et al.}
\graphicspath{{./}{figures/}}

\begin{document}

\title{Magnetic Reconnection During the Post-Impulsive Phase of a Long-Duration Solar Flare: Bi-Directional Outflows as a Cause of Microwave and X-ray Bursts}

\correspondingauthor{Sijie Yu}
\email{sijie.yu@njit.edu}

\author[0000-0003-2872-2614]{Sijie Yu (\cntext{余思捷})}
\affil{Center for Solar-Terrestrial Research, New Jersey Institute of Technology,
323 M L King Jr Blvd, Newark, NJ 07102-1982, USA}

\author[0000-0002-0660-3350]{Bin Chen (\cntext{陈彬})}
\affil{Center for Solar-Terrestrial Research, New Jersey Institute of Technology,
323 M L King Jr Blvd, Newark, NJ 07102-1982, USA}

\author[0000-0002-6903-6832]{Katharine K. Reeves}
\affiliation{Harvard-Smithsonian Center for Astrophysics, 60 Garden St, Cambridge, MA 02138, USA}

\author[0000-0003-2520-8396]{Dale E. Gary}
\affil{Center for Solar-Terrestrial Research, New Jersey Institute of Technology,
323 M L King Jr Blvd, Newark, NJ 07102-1982, USA}

\author[0000-0002-0945-8996]{Sophie Musset}
\affil{SUPA, School of Physics \& Astronomy, University of Glasgow, Glasgow G12 8QQ, UK}
\affil{University of Minnesota, Minneapolis, MN, USA}

\author[0000-0001-5557-2100]{Gregory D. Fleishman}
\affil{Center for Solar-Terrestrial Research, New Jersey Institute of Technology,
323 M L King Jr Blvd, Newark, NJ 07102-1982, USA}

\author[0000-0003-2846-2453]{Gelu M. Nita}
\affil{Center for Solar-Terrestrial Research, New Jersey Institute of Technology,
323 M L King Jr Blvd, Newark, NJ 07102-1982, USA}

\author[0000-0001-7092-2703]{Lindsay Glesener}
\affil{University of Minnesota, Minneapolis, MN, USA}

\begin{abstract}
Magnetic reconnection plays a crucial role in powering solar flares,  production of energetic particles, and plasma heating. However, where the magnetic reconnections occur, how and where the released magnetic energy is transported, and how it is converted to other forms remain unclear. Here we report recurring bi-directional plasma outflows located within a large-scale plasma sheet observed in extreme ultraviolet emission and scattered white light during the post-impulsive gradual phase of the X8.2 solar flare on 2017 September 10. Each of the bi-directional outflows originates in the plasma sheet from a discrete site, identified as a magnetic reconnection site. These reconnection sites reside at very low altitudes ($< 180$ Mm, or 0.26 $R_{\odot}$) above the top of the flare arcade, a distance only $<$3\% of the total length of a plasma sheet that extends to at least 10 $R_{\odot}$. Each arrival of sunward outflows at the looptop region appears to coincide with an impulsive microwave and X-ray burst dominated by a hot source (10--20 MK) at the looptop, which is immediately followed by a nonthermal microwave burst located in the loopleg region. We propose that the reconnection outflows transport the magnetic energy released at localized magnetic reconnection sites outward in the form of kinetic energy flux and/or electromagnetic Poynting flux. The sunward-directed energy flux induces particle acceleration and plasma heating in the post-flare arcades, observed as the hot and nonthermal flare emissions. 

\end{abstract}


\section{Introduction} 
\label{sec:intro}
The most powerful explosive phenomena in the solar system, solar flares accompanied by plasma eruptions, are powered by magnetic energy release in the solar corona facilitated by fast magnetic reconnection. In the standard CSHKP model of eruptive solar flares \citep{1964NASSP..50..451C, 1966Natur.211..695S, 1974SoPh...34..323H,1976SoPh...50...85K}, reconnection occurs in the diffusion region of a reconnection current sheet (RCS) formed in the wake of the eruption of a magnetic flux rope. The latter, when observed by a coronagraph as a large-scale eruptive structure, is referred to as a coronal mass ejection (CME). At both sides of the RCS, magnetized plasma is drawn toward the RCS, where oppositely-directed magnetic fields reconnect, producing bi-directional plasma outflows along the RCS directed away from the reconnection site. Electrons and ions are accelerated at or in the close vicinity of the RCS to high energies \citep{1994Natur.371..495M,2018ApJ...866...62C,Chen2020b,Fleishman2020}. The downward propagating accelerated electrons arrive at the dense chromosphere and drive chromospheric evaporation. Hard X-ray (HXR) emission is produced at the footpoints of the newly reconnected magnetic loops via bremsstrahlung. The dense chromospheric plasma heated to $\sim$10 MK is driven upward by the over-pressure and fills the flare arcade, which produces intense emission in extreme ultraviolet (EUV) and soft X-ray (SXR) wavelengths.

The magnetic reconnection is a driver of the magnetic energy release. The released energy is converted to other forms of energy---bulk flows, heated plasma, and accelerated particles \citep[see, e.g., a review by ][]{2017LRSP...14....2B}. Observational signatures of magnetic reconnection include $X$-shaped \citep{2013NatPh...9..489S, 2015NatCo...6.7598S,2016ApJ...821L..29Z} or $Y$-shaped \citep{2007Sci...318.1591S} magnetic field lines, fan-spine-type structures \citep{2009ApJ...707L..37L,2016ApJ...819L...3Z}, supra-arcade fan or outflows \citep{1999ApJ...519L..93M,2010ApJ...722..329S,2011ApJ...730...98S,2017ApJ...836...55R}, and large-scale thin plasma-sheet-like structures \citep{2011ApJ...727L..52R,2012ApJ...754...13S,2018ApJ...854..122W,2018ApJ...866...64C,2018ApJ...868..148L,2019ApJ...887L..34F,Chen2020b}. However, details of the magnetic reconnection, the associated energy release, and its conversion have yet to be clarified. %In particular, it is unclear, how the magnetic reconnection evolves as the flare progresses from short to long time scales. 

Bi-directional plasma outflows help in probing the magnetic reconnection. The sunward (downward) reconnection outflows usually exhibit as supra-arcade plasma downflows or fast-contracting loops in EUV and SXR data \citep{1996ApJ...459..330F,2008ApJ...675..868R,2013ApJ...767..168L, 2012ApJ...745L...6T,1999ApJ...519L..93M}. Anti-sunward (upward) outflows have also been reported trailing the erupting flux rope \citep{2010ApJ...711.1062N,2017ApJ...841...49C,2018ApJ...866...64C}. Occasionally, both the upward and downward plasma outflows are observed simultaneously \citep{2010ApJ...722..329S,2012ApJ...745L...6T,2013ApJ...767..168L}, which allows pinpointing the reconnection site. A similar pinpointing of the reconnection site with a very high positional accuracy (of $<$1 Mm) has been achieved by \citet{2018ApJ...866...62C} using radio imaging spectroscopy observations of decimetric type III radio bursts. % (emitted by semi-relativistic electron beams). %By tracing a group of semi-relativistic electron beams to their common origin, the site of magnetic reconnection during a solar jet was determined with very high accuracy (within $<$1 Mm).

These plasma outflows carry a significant energy in the form of bulk kinetic energy flux, enthalpy flux, and electromagnetic Poynting flux \citep{2008ApJ...675.1645F,2009ApJ...695.1151B}. To convert these forms of energy  into that of accelerated particles and heated flare plasma, various mechanisms have been proposed involving turbulence or plasma waves \citep{1992ApJ...398..350H,1996ApJ...461..445M,2004ApJ...610..550P,2008ApJ...676..704L,2013ApJ...767..168L,2013MNRAS.429.2515F,Fleishman2020}, fast-mode termination shocks \citep{1986ApJ...305..553F,1994Natur.371..495M,1998ApJ...495L..67T,2012ApJ...753...28G,2013PhRvL.110e1101N,2015ApJ...805..135T,2015Sci...350.1238C,2019ApJ...884...63C,2018ApJ...869..116S,2019ApJ...887L..37K}, collapsing magnetic traps formed by fast-contracting post-reconnection loops \citep{1997ApJ...485..859S,2004A&A...419.1159K,2005ApJ...635..636G, 2006ApJ...647.1472K,2012A&A...546A..85G}, magnetic islands \citep{2006Natur.443..553D,2010ApJ...714..915O}, or Fermi-type acceleration from plasma compression \citep{2018ApJ...866....4L}. 

\begin{figure*}[ht!]
\epsscale{1.2}
\plotone{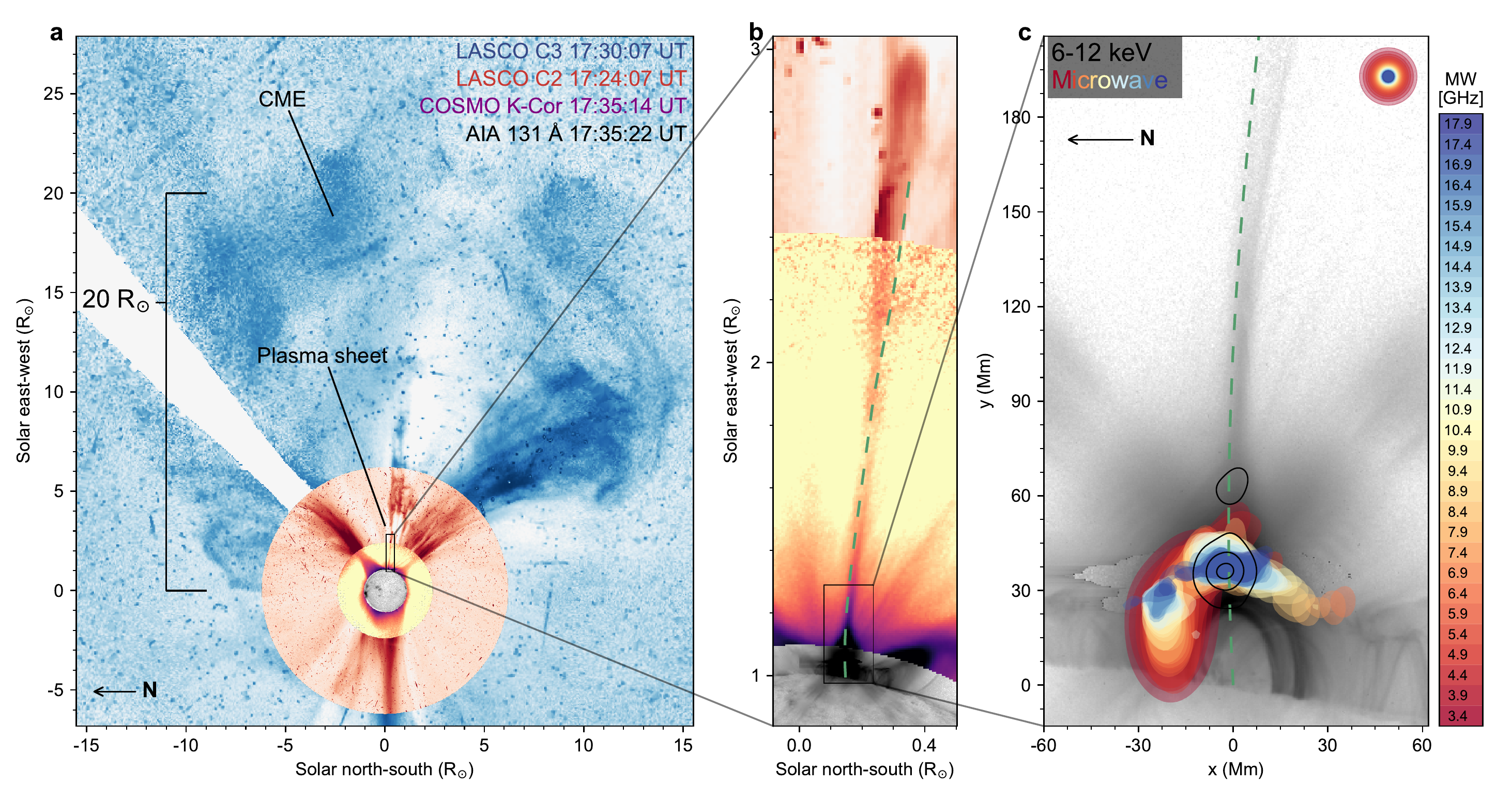}
\caption{(a) Composition of the \textit{SDO}/AIA 131 \AA, \textit{MLSO}/K-cor, \textit{SOHO}/LASCO C2 and C3 white light images, showing the CME bubble and a long plasma sheet connecting the core of CME and the underlying flare site. All images are in reversed grayscale of log intensity, and rotated counter-clockwise by 90$^{\circ}$. (b) Detailed view of the lower portion of the plasma sheet (black box in (a)) seen in EUV and white light. The green dashed curved is used to derive the time-distance maps shown in Fig.\,\ref{fig:outflows}. (c) Further enlarged view of the low-coronal portion of the plasma sheet and the flare arcade (black box in (b)). Note that for the ease of further discussions, we have set the origin (i.e., $x=0$ Mm and $y=0$ Mm) at the limb location immediately below the plasma sheet with a helioprojective longitude and latitude coordinate of (948$"$, -140$"$). The \textit{EOVSA} microwave emission at 30 spectral windows is displayed as filled contours (25\% of the respective maximum intensity), color-coded in frequency according to the colorbar. The filled circles on the upper right corner represent the full-width-half-maximum size of the restoring beams at the respective frequencies. {\it RHESSI} 6-12 keV X-ray source is superposed as black contours (10\%, 50\%, 90\% of the maximum). \label{fig:flare}}
\end{figure*}

Occasionally, HXR (and sometimes $\gamma$-ray) sources are observed at or above the top of bright EUV/SXR flare arcades \citep{1994Natur.371..495M, 2006A&A...456..751B,2006AA...446..675V,2008ApJ...678L..63K,2008A&ARv..16..155K,2010ApJ...714.1108K,2012ApJ...748...33C,2013A&A...551A.135S,2013ApJ...767..168L,2014ApJ...780..107K,2018ApJ...867...82D,2018ApJ...865...99P}. They have been regarded as an  evidence of electron acceleration  in the above-the-looptop (ALT) region \citep[e.g.,][]{2006A&A...456..751B,2010ApJ...714.1108K,2013ApJ...777...33C,2013ApJ...767..168L}. Recently, microwave imaging spectroscopy data from the Expanded Owens Valley Solar Array (EOVSA) have offered unprecedented spatially- and temporally-resolved measurements of nonthermal electrons and magnetic field in the flaring region \citep{2018ApJ...863...83G}.  \citet{Fleishman2020} reported a fast decay of magnetic field in the ALT region.  \citet{Chen2020b} found that this region coincides with a local minimum of magnetic field strength, referred to as a ``magnetic bottle'', where microwave-emitting, mildly relativistic electrons are  concentrated. These observations imply that the ALT region, where plasma outflows interact  with the underlying flare arcade, plays a crucial role in magnetic energy release and electron acceleration. Reported temporal correlation between the outflows and impulsive X-ray and/or radio emission \citep{2004ApJ...605L..77A,2007A&A...464..735K,2010ApJ...711.1062N,2015Sci...350.1238C,2016ApJ...828..103T} further supports the role of the plasma outflows.

In this study, we report well-connected signatures of magnetic reconnection, plasma heating, and electron acceleration observed during the post-impulsive gradual phase of the X8.2-class eruptive solar flare on 2017 September 10. The combined white light and EUV imaging observations allow us to identify the timing and location of multiple intermittent reconnection events by bi-directional plasma outflows in an extremely long plasma sheet. The arrivals of the plasma downflows at the looptop correlate with plasma heating events that manifest as impulsive X-ray bursts. In the meantime, nonthermal microwave bursts, obtained by EOVSA, are detected in the loopleg region, which have no response in {\it RHESSI} hard X-rays. Such a chain of reconnection-associated observational signatures offers a new view of the energy release and conversion processes with a level of clarity not previously achieved.

In Section 2.1, we present imaging spectroscopy observations of impulsive microwave and X-ray bursts in the flare looptop and arcade (Section 2.2). In Section 2.3, we examine the microwave and X-ray bursts using spectral analysis. In Section 2.4, we present the detection of multitudes of bi-directional plasma outflows using white light and EUV imaging data that appear to correlate with the microwave and X-ray bursts. We discuss the implications of the observational results in Section 3, particularly on the role of the plasma outflows in energy transport and conversion.

\section{Observations} 
\label{sec:obs-res}

The long-duration X8.2-class eruptive flare occurred close to the west solar limb on 2017 September 10. The event is associated with a fast white-light coronal mass ejection (CME), which has a speed of $>$ 4000 km $\rm{s^{-1}}$ \citep{2018ApJ...863L..39G} and is accompanied with a type II radio burst \citep{2019NatAs...3..452M}. The eruption is well observed in microwaves by \textit{EOVSA} \citep{2018ApJ...863...83G} in 2.5--18 GHz, in EUV by the Atmospheric Imaging Assembly on board the Solar Dynamics Observatory (\textit{SDO}/AIA; \citealt{2012SoPh..275...17L}) and Solar Ultraviolet Imager on board the NOAA GOES-R satellite \citep{2018ApJ...852L...9S}, in white-light by the COSMO K-Coronagraph of the Mauna Loa Solar Observatory (\textit{MLSO/K-Cor}; \citealt{2003SPIE.4843...66E}) and the Large Angle and Spectrometric Coronagraph Experiment on board the solar and heliospheric observatory ({\it SoHO}/LASCO; \citealt{1995SoPh..162..357B}), and in X-Rays by Reuven Ramaty High-Energy Solar Spectroscopic Imager (\textit{RHESSI}; \citealt{2002SoPh..210....3L}) and the
Gamma-ray Burst Monitor aboard the \textit{Fermi} spacecraft (Fermi/GBM;  \citealt{2009ApJ...702..791M}). 

\begin{figure*}[ht!]
\epsscale{1.0}
\plotone{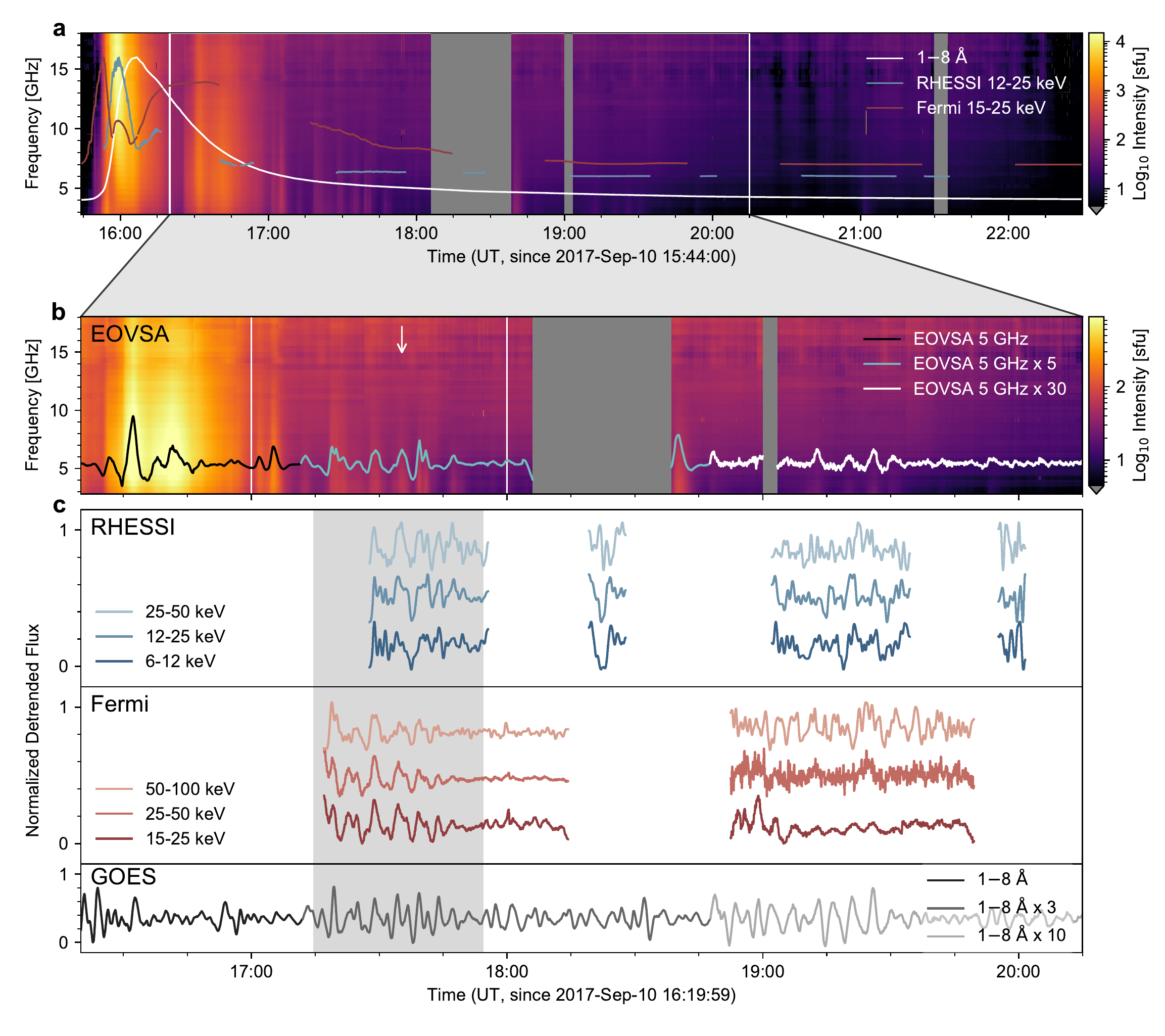}
\caption{(a) \textit{EOVSA} total-power (full-disk-integrated) microwave dynamic spectrum of the entire event in 2.5--18 GHz. Color presents the flux density in sfu. Overplotted are normalized time profiles of {\it GOES} 1-8 \AA\ SXR (white curves), {\it RHESSI} 12--25 keV and {\it Fermi} 15--25 keV X-ray counts (solid blue and red curves, respectively). (b) Enlarged \textit{EOVSA} total-power dynamic spectrum of the post-impulsive phase. The time window is demarcated by a pair of vertical lines in (a). Overplotted is the \textit{EOVSA} 5 GHz light curve after removing the slowly-varying background. In order to aid visual comparison, the 5 GHz time profile between 17:12 and 18:47 UT and after 18:47 UT is multiplied by a factor of 5 and 30, respectively (shown in distinct colors). (c) Detrended light curves of {\it RHESSI} 6--50 keV and {\it Fermi} 25--50 keV X-ray counts, and detrended {\it GOES} 1-8 \AA\ flux. The detrended {\it GOES} after 17:12 UT is amplified in a similar fashion to the \textit{EOVSA} 5.0 GHz time profile in (b). Each time profile is elevated progressively for visualization purpose.} \label{fig:timeprofile}
\end{figure*}

The event began at $\sim$15:35 UT when a pre-existing filament started to erupt. After $\sim$15:50 UT, the erupting filament developed into a teardrop-shaped dark cavity as seen in SDO/AIA images \citep{2018ApJ...853L..18Y,2018ApJ...868..107V} and microwaves \citep{Chen2020a}. The acceleration of the flux rope peaked at around 15:54 UT \citep{2018ApJ...868..107V}, which coincided with an early impulsive peak in microwaves and HXRs \citep{Chen2020a}. The main peak of the impulsive microwave/HXR emission occurred at around 16:00 UT \citep{2018ApJ...863...83G,Fleishman2020}. By 17:30 UT well into the decay phase of the event, the CME front has already propagated to more than 20 solar radii (R$_\odot$) above the solar surface (Figure\,\ref{fig:flare}(a)). A long and thin plasma sheet structure is present in the wake of the CME above the flare arcade (Figure\,\ref{fig:flare}; see also \citealt{2018ApJ...853L..18Y,2018ApJ...854..122W,2018ApJ...868..148L,2019ApJ...887L..34F,Chen2020b}), which extends into the \textit{SOHO}/LASCO C3 field of view with a total length of at least 8 $R_{\odot}$ \citep{2018ApJ...866...64C,lee_formation_2020}. A \textit{RHESSI} 6--12 keV X-ray source is present at the looptop region (open contours in Fig.\,\ref{fig:flare}c). No footpoint X-ray source is detected at this time.

\begin{figure*}[ht!]
\epsscale{1.2}
\plotone{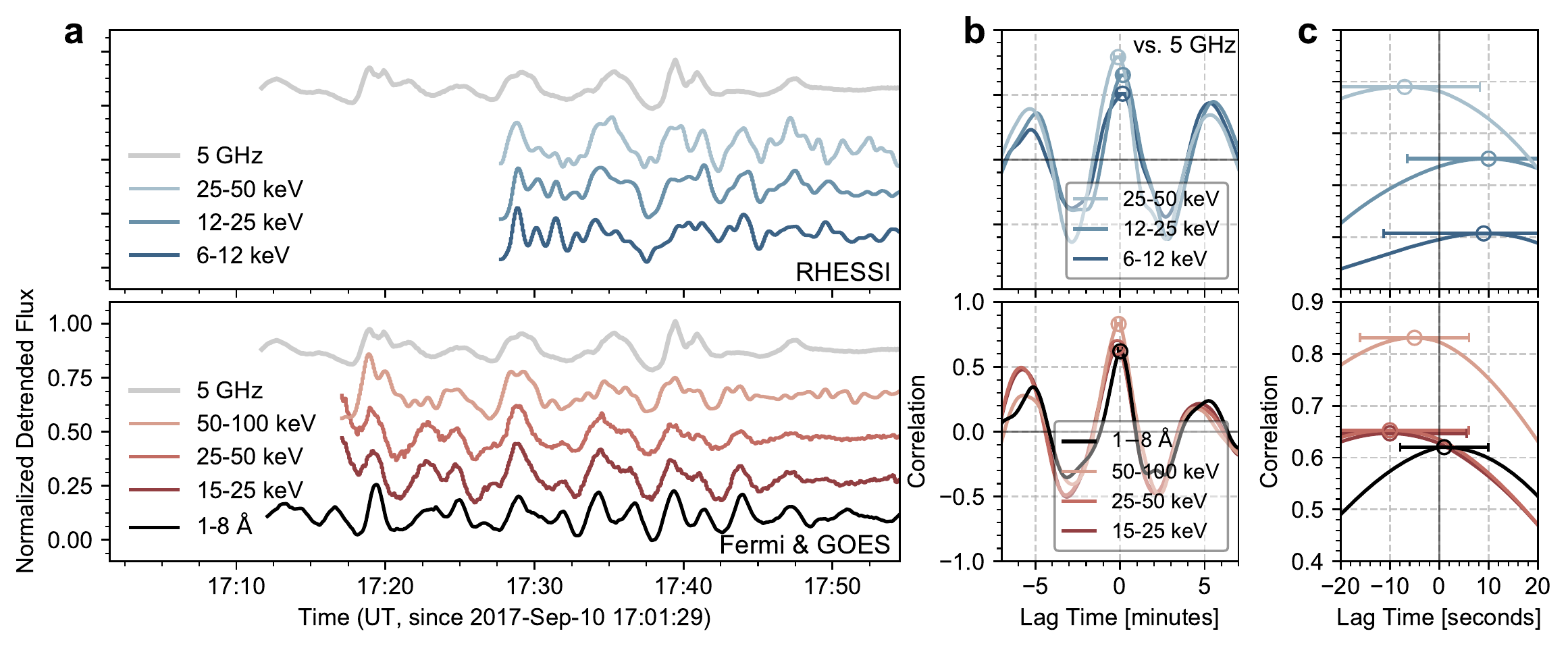}
\caption{(a) Thirty-minute details of the detrended X-ray light curves from \textit{RHESSI} (upper) and \textit{Fermi} \& \textit{GOES} (lower). The time window is demarcated by the gray shaded area in Fig. \ref{fig:timeprofile} (c). Overplotted are the \textit{EOVSA} 5 GHz light curves (gray curve) for comparison purpose. (b) Cross-correlation functions of the detrended X-ray light curves with the 5.0 GHz light curve. The color scheme is same as that used for the light curves in (a). (c) Blowup of the cross-correlation functions near the peak regions. The peaks are shown as hollow circles with error bars. (Negative/Positive peak location denotes X-ray is ahead/behind of microwave.)} \label{fig:timecorr}
\end{figure*}

\subsection{Impulsive Microwave and X-ray Bursts}
\label{sec:obs-res:flare}
An overview of the long-duration event based on \textit{EOVSA} and \textit{RHESSI} data was provided by \citet{2018ApJ...863...83G}. More detailed studies of the flux rope morphology, magnetic field variation along the RCS feature, and magnetic field decay during the initial and main impulsive phase have been reported in our earlier publications \citep{Chen2020a,Chen2020b,Fleishman2020}. In this study, we focus on the post-impulsive, long-duration gradual phase between 16:20 and 20:15 UT (Fig.\,\ref{fig:timeprofile}(b)), shortly after the main impulsive phase at $\sim$16:00 UT (c.f., Fig.\,\ref{fig:timeprofile}(a), when the brightest microwave emission of over 10,000 solar flux unit, or sfu, is present). During this period, the \textit{EOVSA} total-power (full-disk integrated) microwave dynamic spectrum contains multiple broadband bursts. These bursts have an impulsive appearance in the dynamic spectrum and light curves. The relative amplitudes of the bursts at 5 GHz, represented as the ratio of the peak brightness of each burst to the pre- and post-burst background $I_{\rm pk}/I_{\rm bkg}$, is between 1.5\%--30\% (solid curves in Fig.\,\ref{fig:timeprofile}(b)). These bursts have an average duration of $\sim$4 minutes and an average recurrence period of $\sim$5.6 minutes. The individual microwave bursts correlate with weak X-ray bursts at 6--50 keV observed by \textit{RHESSI} and at 15-100 keV by \textit{Fermi/GBM}. The X-ray bursts, however, have very small amplitudes of only a few percent and can only be distinguished in the detrended light curves, after the slow-varying background has been removed (Fig.\,\ref{fig:timeprofile}(c)). Some of the microwave bursts correspond to quasi-periodic impulsive peaks in the detrended \textit{GOES} 1--8 \AA\ SXR light curve after the slow-varying back-ground has been removed (Fig.\,\ref{fig:timeprofile}(c)). The latter was reported in a recent paper by \citet{2019ApJ...875...33H}, who attributed these quasi-periodic features with a period of $\sim$150 seconds to magnetohydrodynamic (MHD) oscillations in the post-flare arcade. 

To explore the temporal relation of the bursts seen in the X-rays and microwave quantitatively, we perform cross-correlation analysis between the detrended X-ray light curves obtained by \textit{RHESSI}, \textit{Fermi}, and \textit{GOES} at a variety of energy ranges, and the detrended 5.0 GHz EOVSA light curve in 17:15--17:50 UT (Fig.\,\ref{fig:timecorr}a). Figure\,\ref{fig:timecorr}(b) shows that the X-ray and microwave light curves are well correlated with a high correlation coefficient of 0.62--0.83. No time lag is found between the bursts seen in the X-ray and microwave within the uncertainty of the cross-correlation peak location. The latter is determined by $\sigma\approx0.75W_c/(1+h/\sigma_\mathrm{n}$), where $\sigma_\mathrm{n}$ is the noise level in the cross-correlation function, and $h$ and $W_c$ are the height and the half-width at half-maximum of the peak in the function, respectively \citep{1979AJ.....84.1511T,1987ApJS...65....1G}. In Fig.\,\ref{fig:timecorr}(c), we show that all the peaks of the cross-correlation coefficients are in the range of $-$10 s to 10 s, which fall within the $\pm\sigma$ uncertainties (horizontal error bars).
% It is noted that the correlation coefficient between the high energy X-rays with microwave tends to be higher than their counterparts at lower energies.

\subsection{X-ray spectroscopy and imaging}
\label{sec:obs-res:X-ray}

%  The looptop source underwent a gradual ascent (see also \citealt{2019ApJ...875...33H}).
\begin{figure}[ht!]
\epsscale{1.2}
\plotone{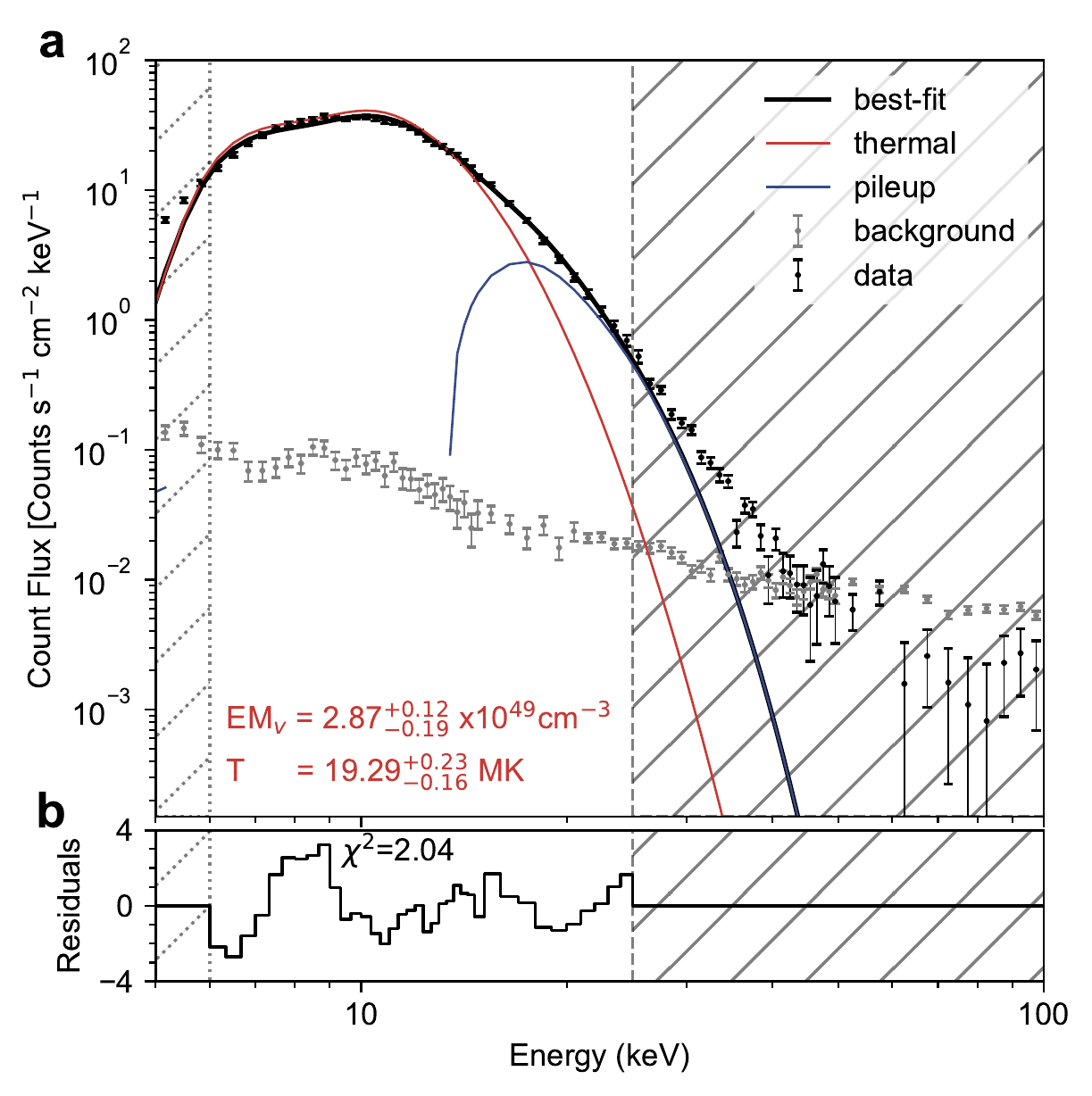}
\caption{Example {\it RHESSI} HXR count spectrum integrated between 17:34:44 UT and 17:35:12 UT. (a) The background-subtracted count spectrum is shown in black symbols and the background is shown in gray symbols. The hatched area indicates the energy ranges where the observed X-ray count flux is excluded from the spectral fitting. The fitted thermal component, pileup component and the best-fit spectrum (sum of components) are shown in red, blue, and black curves, respectively. Also shown are the best-fit temperature $\mathrm{T}$ and volume emission measure $\mathrm{EM_V}$. The quoted errors in $\mathrm{T}$ and $\mathrm{EM_V}$ denote a 99\% confidence range with the fit determined by Monte Carlo analysis. (b) Normalized residuals vs. energy in units of the standard deviation of the count statistics. \label{fig:hxrSpectraFitting}}
\end{figure}

We analyze the \textit{RHESSI} full-disk-integrated X-ray spectrum during the peak of a selected burst at around 17:35 UT integrated over a 28-s interval (indicated by a white arrow in Figure\,\ref{fig:timeprofile}(b)) using the standard X-ray spectral analysis tool {\tt OSPEX} available in the SolarSoft Ware \citep[SSW;][]{1998SoPh..182..497F} IDL package. \citet{2019RAA....19..173N} reported a significant pulse pileup effect on the \textit{RHESSI} spectra in this event. Pileup results when two lower-energy photons arrive at the detector within a short time and are counted by the detector as a single higher-energy photon (\citealt{2002SoPh..210...33S}). We check the impact of the pulse pileup effect at $\sim$17:35 UT using the standard routine {\tt hsi\_pileup\_check} in SSW. We find while pileup strongly affects the spectrum above $\sim$20 keV, the lower-energy range is almost pileup-free. Such pileup effect on relatively higher energies is also implied by the similar temporal behavior of the light curves between \textit{RHESSI}/\textit{Fermi} 6--50 keV and \textit{GOES} detrended 1--8 \AA\ SXR (as mentioned in section \ref{sec:obs-res:flare}), which is consistent with a dominant thermal plasma at lower energies and some pileup at higher energies. We restrict our spectral fit to only the low-energy range (6--25 keV). We include the pileup in the detector response matrix for spectral fitting by adding the pileup module {\tt pileup\_mod} as a fitting component. Figure\,\ref{fig:hxrSpectraFitting} shows the observed X-ray count flux spectrum and the spectral fit result that corresponds to an isothermal model with a temperature of $\mathrm{T}\approx19\,\mathrm{MK}$ and a volume emission measure $\mathrm{EM_V}\approx2.9\times 10^{49}\,\mathrm{cm}^{-3}$ (red curve), together with pulse pileup correction (blue curve). We note that the uneven pattern in the residuals (particularly those in 6--9 keV; Figure~\ref{fig:hxrSpectraFitting}(b)) may be an indication of unaccounted contributions from, e.g., the emission line complexes at 6.65 keV (Fe) and 8 keV (Fe/Ni) or imperfect pileup corrections. We explore the uncertainties in the best-fit parameters by employing the Monte Carlo analysis implemented in {\tt ospex}. The estimated uncertainties in the fit results $\mathrm{T}$ and $\mathrm{EM_V}$ are relatively small (shown in Figure\,\ref{fig:hxrSpectraFitting}). We caution that, however, these uncertainties should be considered as lower limits due to the simplifications made in the model (isothermal continuum) as noted earlier. Nevertheless, the X-ray spectrum below $\sim$20 keV favors a thermal source of $\sim$19 MK.
% Although the fit is subject to uncertainties noted above, we conclude that the X-ray spectrum below $\sim$20 keV is consistent with thermal emission from $\sim$20 MK plasma. 
While the pileup effect renders the spectral analysis difficult for the energy range above $\sim$20 keV, there is a hint for the possible presence of a weak nonthermal component in the $\sim$30--40 keV range, where the observed HXR counts exceed the pileup component and the background.

% \textbf{While the pileup effect at higher energies hinders us from constraining a possible nonthermal component based on the spectral fit, we cannot rule out the presence of a weak nonthermal component at the looptop region. In figure\,\ref{fig:timecorr}(b) and (c), the cross-correlation analysis shows that a tendency that the X-ray light curves at higher energies has a higher similarity with microwave 5 GHz light curve. The tendency is likely caused by the subtle, common features present in both higher energies X-ray light curves and the microwave 5 GHz light curves, which are largely absent or deformed in their counterparts at lower energies that dominated by thermal emission and pileup. Two examples of such features in the \textit{Fermi} 50--100 keV light curves are highlighted by the black arrows in Figure\,\ref{fig:timecorr}(a). These features can not arise from pileup effect --- pileup tend to make X-ray emission at higher energies resembling those at lower energies, but can not make the former differing from the latter. Therefore, the tendency implies the likely presence of a weak nonthermal component in the X-ray spectrum as well.}

\begin{figure}[ht!]
\epsscale{1.2}
\plotone{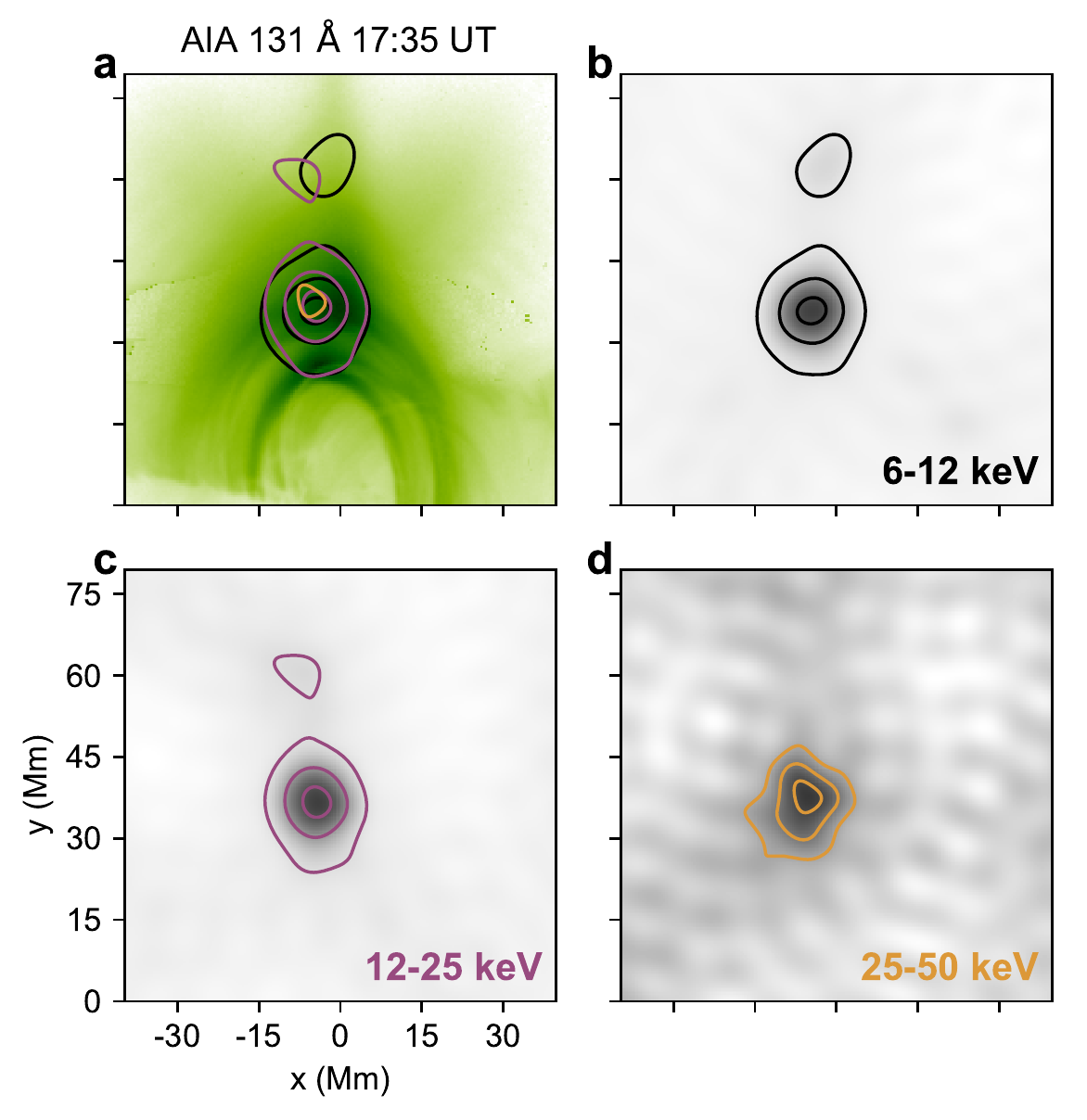}
\caption{An example of \textit{RHESSI} HXR images during the post-impulsive phase. (a) Contours of the 6--12 keV (black), 12--25 keV (green) and 25--50 keV (purple) sources on \textit{SDO}/AIA 131 \AA\ image at 17:35 UT with inverted color scale. The contour levels are 10\%, 50\% and 90\% of the maximum for the 6--12 keV and 12--25 keV bands, and 90\% of the maximum for the 25--50 keV band. (b-d) Same {\it RHESSI} images as in (a) shown separately. Note the 25--50 keV image is strongly affected by pulse pileup.} 
\label{fig:hxr-images}
\end{figure}

We reconstruct {\it RHESSI} X-ray images using the {\tt CLEAN} algorithm \citep{2002SoPh..210...61H} with measurements from detectors 3, 6 and 8. Time-series images were made in three energy bands, 6--12 keV, 12--25 keV, and 25--50 keV, over two {\it RHESSI} observing windows in 17:28--17:55\,UT and 19:02--19:35\,UT with an integration time of 60 s for each individual image. Figure\,\ref{fig:hxr-images} shows an example of {\it RHESSI} images at the three energy bands. There is a persistent X-ray looptop source in 6--12 keV and 12--25 keV located near the apex of the flare arcade. Spectral analysis described above suggests that this source is dominated by thermal emission from a hot source with a temperature $T\approx 19$ MK. The higher-energy 25--50 keV X-ray source appears co-spatial with the lower-energy source (Fig.\,\ref{fig:hxr-images}(d)). However, this band is severely affected by the pileup effect (c.f., Fig.\,\ref{fig:hxrSpectraFitting}). Thus we choose not to perform an in-depth analysis on the images of this band. %nonthermal looptop source above 25 keV are nearly cospatial within a few Mm with the thermal source. Since the looptop source position are similar across the three energy bands, we use the image in 6-12 keV for further analysis. 

In the \textit{RHESSI} 6--12 keV and 12--25 keV images, a weaker secondary coronal source is present above the primary source by $\sim$27 Mm. The secondary source is located near the tip of the cusp-shaped flare arcade, seen by \textit{SDO}/AIA 131 \AA\ (Fig.\,\ref{fig:hxr-images}(a)), and is persistent over nearly the entire time of interest. Although its higher-energy counterpart is elusive due to the pileup effect, this secondary source appears reminiscent of the ALT HXR sources seen in the ``Masuda-type'' flares, where a HXR source is located slightly above the bright SXR/EUV flare arcade \citep{1994Natur.371..495M,2006AA...446..675V,2013ApJ...767..168L,2014ApJ...780..107K}.
We estimate the full width at half-maximum (FWHM) size of the primary and secondary source as $\sim$11 Mm and $\sim$9 Mm, respectively. Since the primary looptop source dominates the X-ray flux, we adopt its source area to estimate an average column emission measure of $\mathrm{EM_C}\approx 9.3\times 10^{30}\,\mathrm{cm}^{-5}$ in the looptop source.
% \footnote{
It should be noted that the source area in the {\tt CLEAN} images characterized by its full-width-half-maximum size may result in an  overestimation \citep{2009ApJ...698.2131D,2010ApJ...717..250K} and should be treated as the upper limit. However, the true source size should not be more than 10--20\% smaller, and will not strongly affect our order-of-magnitude estimate of $\mathrm{EM_C}$ here.
% }. 

\subsection{Microwave imaging spectroscopy} 
\label{sec:obs-res:microwave}

\begin{figure*}[ht!]
\epsscale{1.2}
\plotone{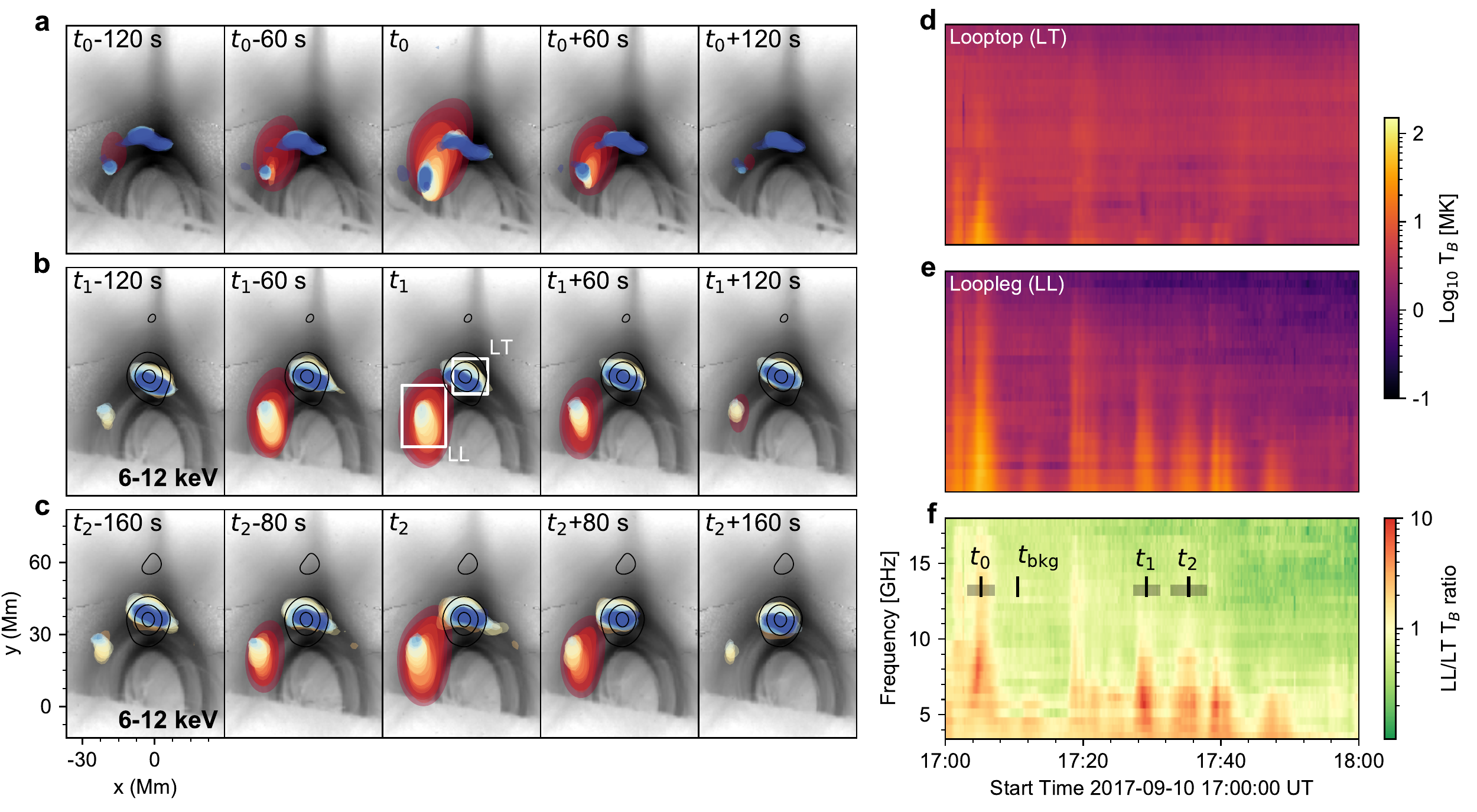}
\caption{Spatial evolution of the microwave bursts observed by \textit{EOVSA}. (a--c) Image sequences of the microwave emission (filled contours at 45\% of the maximum brightness temperature of all images in the selected time interval; color represents different frequencies with same hue as in Figure\,\ref{fig:flare}(c)) overlaid on \textit{SDO}/AIA 131 \AA\ images during three time intervals. {\it RHESSI} 6-12 keV 15\%, 50\%, 90\% contours are also superposed as black contours. The corresponding time $t_0$, $t_1$ and $t_2$, and the relevant time intervals are marked by the black vertical bars and the shaded stripes, respectively, in panel (f). (d) and (e) EOVSA \textit{spatially-resolved} (vector) dynamic spectra derived from looptop and loopleg source from 17:00 UT to 18:00 UT, as well as their intensity ratio (f). The time range of the dynamic spectra is demarcated by the vertical dashed lines in Figure\,\ref{fig:timeprofile}b. \label{fig:mw_motion}}
\end{figure*}

\textit{EOVSA} provides microwave images in 2.5--18 GHz with 134 spectral channels spread over 31 spectral windows. In this work we combine each of the upper 30 spectral windows centered from 3.4 to 17.9 GHz to produce images at 30 equally spaced frequencies. Figure\,\ref{fig:flare}(c) shows the microwave images at these 30 frequencies at 17:35 UT (filled contours). The overall morphology of the evolving microwave source is consistent with the shape and orientation of the EUV flare arcade. At high frequencies (blue colors), there are two distinct sources: one coincides with the looptop HXR source, while the other is in the northern leg of the flare arcade. At low frequencies, the microwave source concentrates in the northern leg of the flare arcade (on the left side in Fig.\,\ref{fig:flare}(c)), where the second (weaker) high-frequency microwave source is located. The microwave emission at all frequencies is weak or absent in the southern leg of the flare arcade (right side in the diagram).

The time-series of microwave images reveals that the impulsive component of the microwave emission in each burst is mainly from the loopleg source. Fig.\,\ref{fig:mw_motion}(a--c) shows such time-sequence images for three selected microwave bursts peaking at 17:05 UT, 17:29 UT, and 17:35 UT, respectively (denoted as $t_0$, $t_1$, and $t_2$ in Fig.\,\ref{fig:mw_motion}(f)). In these time-sequence images, the loopleg source shows a large variation in intensity during each burst (by a factor of up to 15). In contrast, the looptop source appears relatively stable with more-minor variations in morphology and intensity.

The dominance of loopleg brightening is better demonstrated in the \textit{spatially-resolved}, or ``vector'', dynamic spectra shown in Figs. \ref{fig:mw_motion}(d--f). First introduced by \citet{2015Sci...350.1238C} using dynamic imaging spectroscopy data from the Karl G. Jansky Very Large Array (VLA), the technique of vector dynamic spectra takes advantage of the spatially, spectrally, and temporally resolved data to derive a radio dynamic spectrum for each selected region of interest in the spatial domain. This spatial separation allows the study of the temporal and spectral properties intrinsic to each radio source of interest. Here we select the looptop and loopleg sources (shown as the white boxes in Fig.\,\ref{fig:mw_motion}(a)) and derive the maximum brightness temperature $T_B$ within each box as a function of time and frequency. The resulting vector dynamic spectra for the loopleg and looptop source are shown in Fig.\,\ref{fig:mw_motion}(d) and (e), respectively, and their intensity ratio is shown in Fig.\,\ref{fig:mw_motion}(f).  Although the loopleg and looptop source both display impulsive features, the bursts in the loopleg source are up to 10 times stronger than the looptop source (c.f., ratio spectrum in Fig.\,\ref{fig:mw_motion}(f)).

\begin{figure*}[ht!]
\epsscale{1.2}
\plotone{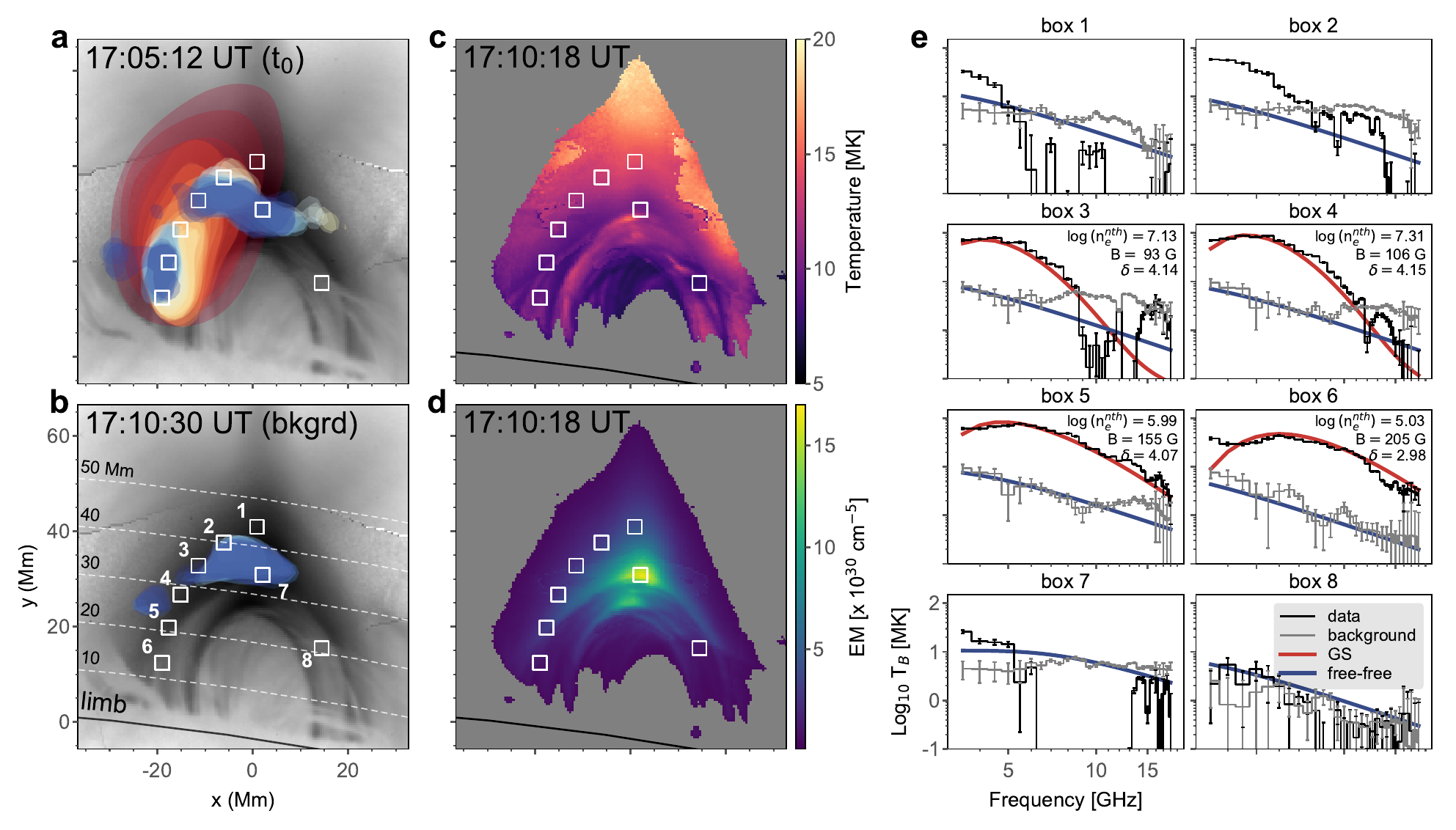}
\caption{Spatially resolved microwave spectra at various points in the flare arcade. Enlarged view of the microwave burst (a) at $t_0$ and (b) at a non-bursting time $t_{\rm bkg}$ selected as the background (filled contours at 30\% of the maximum brightness temperature of all images at $t_0$; same color scheme as in Figure\,\ref{fig:mw_motion}(a)). (c) EM-weighted temperature and (d) integrated column emission measure $\mathrm{EM_C}$ maps obtained at $t_{\rm bkg}$ over the temperature range of 0.5--30 MK. (e) microwave brightness temperature spectra $T_{B}(\nu)$ of the burst peak at $t_0$ from eight selected locations (marked as numbered white boxes in (b)). The background-subtracted burst spectra and their corresponding background spectra derived at $t_{bkg}$ are shown as black and gray histograms, respectively, with error bars. The observed  emission at $t_{\rm bkg}$ is well matched by thermal free-free microwave spectra (blue lines) calculated using the AIA-derived DEM, while the burst enhancements in the loopleg source (boxes 3-6) are well fit by nonthermal gyrosynchrotron spectra (red lines) with the parameters shown in each plot. \label{fig:mwSpec}}
\end{figure*}

%\subsection{Signatures of energetic electrons}  
%\label{sec:obs-res:mw_diag}
%These microwave bursts appear to be produced by energetic electrons presented in the loop-leg region.
To investigate the looptop and loopleg emission, we derive the spatially-resolved microwave brightness temperature spectra $T_{B}(\nu)$ obtained at different spatial locations. Figure\,\ref{fig:mwSpec}(e) shows the background-subtracted $T_{B}(\nu)$ spectra obtained from several selected locations along the flare arcade (white boxes in Fig.\,\ref{fig:mwSpec}(a--d)) at 17:05 UT ($t_0$ in Fig.\,\ref{fig:mw_motion}), during the peak of one of the brightest microwave bursts (solid black curves in Fig.\,\ref{fig:mwSpec}(e)). The background spectra $T_B^{\rm bk}(\nu)$, shown as gray curves in Fig.\,\ref{fig:mwSpec}(e), are obtained at a time just after this microwave burst, denoted as $t_{bkg}$ in Fig.\,\ref{fig:mw_motion}(f). At the lowest frequencies, the background $T_B$ spectra follow a power law with a slope close to $-2$, suggesting an optically-thin, bremsstrahlung origin \citep[e.g.,][Section 10.2]{FT_2013}. At higher frequencies, the spectra are more complex with a flat or rising $T_B$ toward higher frequencies, which is suggestive of a nonthermal gyrosynchrotron contribution.

%For the purpose of analyzing the net contribution during the microwave bursts, in 
Fig.\,\ref{fig:mwSpec}(e) shows the background-subtracted microwave spectra as solid black curves. The spectra in the loopleg region (boxes 3-6) show characteristics of gyrosynchrotron radiation due to nonthermal electrons gyrating in the coronal magnetic field \citep[e.g.,][]{1985ARA&A..23..169D}. We adopt the method described in \citet{Fleishman2020} to fit the spectra using a gyrosynchrotron emission model from an isotropic and homogeneous nonthermal electron source with a power-law energy distribution. From the spectral fit, we obtain the magnetic field strength $B$, the power-law index of the electron energy distribution $\delta$, and the total number density of nonthermal electrons $n^{\mathrm{nth}}_{\mathrm{e}}$ integrated above 100 keV. The robustness and confidence level of the fit parameters are evaluated using a Markov chain Monte Carlo (MCMC) method, described in detail in \citet{Chen2020b}. The derived magnetic field strength increases from $93\,\mathrm{G}$ to $205\,\mathrm{G}$ from box 3 to box 6, corresponding to height range from $\sim$35 Mm to $\sim$15 Mm. This is consistent with an expected increase of the magnetic field strength in a coronal loop towards lower heights. The values are also consistent with spectropolarimetric measurements of the magnetically sensitive Ca II 8542 \AA\ line in the same post-arcade region about one hour before our time of interest \citep{2019ApJ...874..126K}. The power-law index of the electron energy distribution $\delta$, which ranges from 3.0--4.2, indicates a hardening toward lower heights (i.e., toward the loopleg region). The total number density of the nonthermal electrons $n^{\mathrm{nth}}_{\mathrm{e}}$ above 100 keV is $\sim 10^5$--$10^7\,\mathrm{cm}^{-3}$, which is a small fraction of the thermal electron density in the same region (of order $10^{10}\,\mathrm{cm}^{-3}$, estimated based on the column emission measure shown in Fig.\,\ref{fig:mwSpec}(d) and an assumed column depth of a few tens of Mm). 

In the looptop region, however, the microwave spectra (except the lowest few frequencies) show very little increment above the post-burst background (boxes 1, 2, and 7). The spectra (without background subtraction) are consistent with free-free (bremsstrahlung) radiation from hot thermal plasma with a temperature of 10--15 MK and column emission measure of $2.9\text{--}15.0\times10^{30}$ cm$^{-5}$ (blue curves in Fig.\,\ref{fig:mwSpec}(e)). The latter is constrained using a differential emission measure (DEM) analysis method \texttt{xrt\_dem\_iterative2} \citep{Weber2004, Golub2004} based on imaging data at six \textit{SDO}/AIA EUV passbands (94 \AA, 131 \AA, 171 \AA, 193 \AA, 211 \AA, and 335 \AA). The microwave spectral and AIA DEM analysis of the looptop source is consistent with that derived from {\it RHESSI} (temperature of $19\,\mathrm{MK}$ and column emission measure of $9.3\times 10^{30}\,\mathrm{cm}^{-5}$; See section \ref{sec:obs-res:X-ray}).

% \textbf{How does the spectral analysis results of the looptop source compare with RHESSI spectral analysis?}
% box 1 Temperature: 15.3 MK, EM: 3.79e+30 cm-5, nele: 5.09e+10 cm-3 if depth is 20.0 arcsec
% box 2 Temperature: 14.2 MK, EM: 3.28e+30 cm-5, nele: 4.74e+10 cm-3 if depth is 20.0 arcsec
% box 7 Temperature: 10.6 MK, EM: 1.56e+31 cm-5, nele: 1.03e+11 cm-3 if depth is 20.0 arcsec

\begin{figure*}[ht!]
\epsscale{1.1}
\plotone{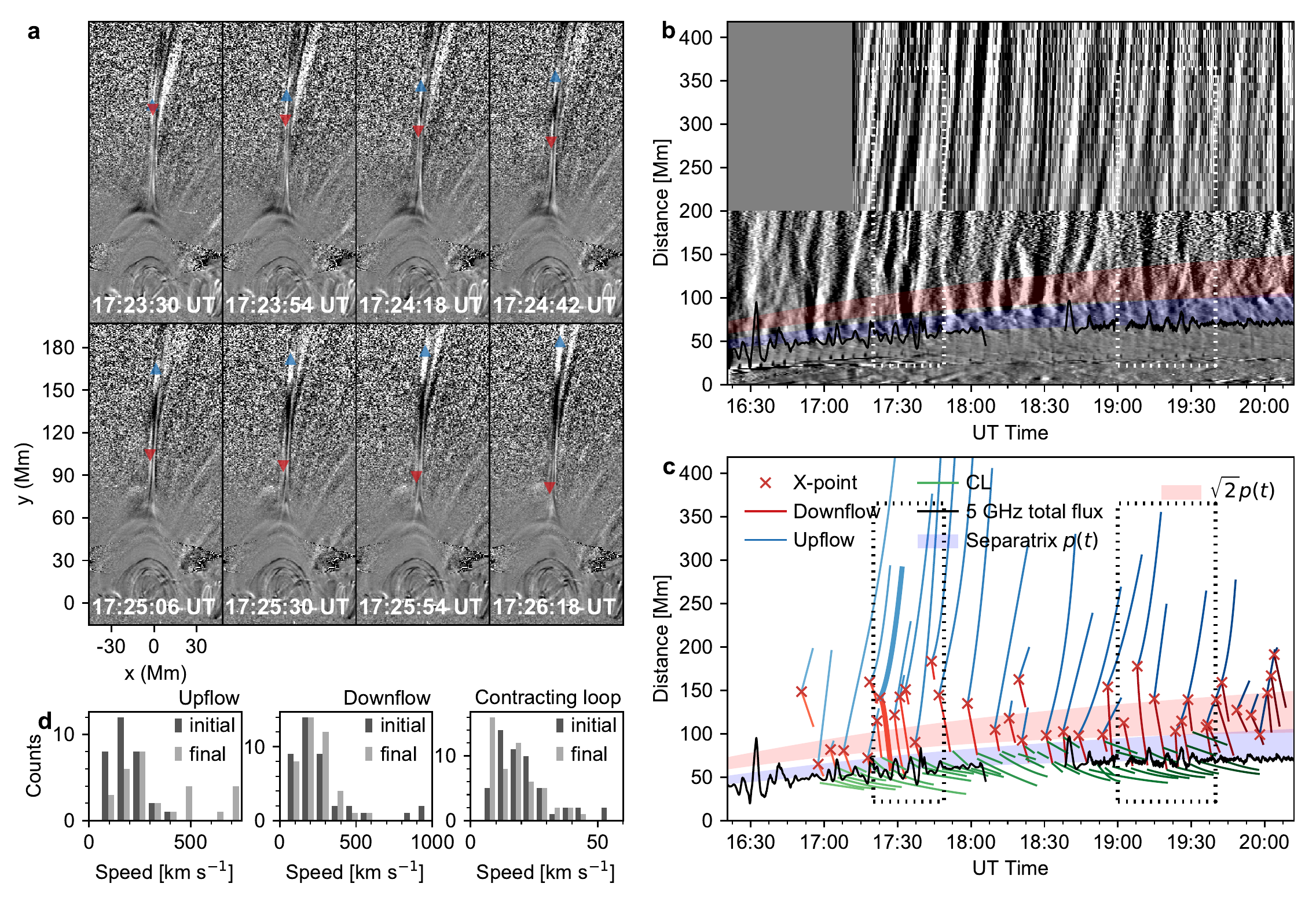}
\caption{(a) Successive \textit{SDO}/AIA 131 \AA\ background-detrended images that show bi-directional outflows diverging from a compact region. The upward- and downward-moving EUV outflows are marked by red and blue triangles, respectively. Their tracks in the time--distance plot are highlighted as a pair of thick blue and red curves in (c). (b) Composite time-distance plot of {\it MLSO}/K-cor white light and \textit{SDO}/AIA 131 \AA\ background-detrended images at a cut made along the RCS (green dashed curve in Figure\,\ref{fig:flare}(b) and (c)). The upper edge of the AIA field of view is at $\sim$ 200 Mm. The blue shaded region marks the separatrix region, denoted by $p(t)$, that divides the fast plasma downflows and the slow contracting loops. The red shaded region shows the predicted location of the reconnection X point at a height of $\sqrt{2}p(t)$, according to the idealized 2D flare model in \citet{2018ApJ...858...70F}. (c) Tracks of the fast upward/downward outflows (blue/red) and slow contracting loops (CL; green) in the time-distance plot. The inferred X points from the diverging sites of the bi-directional outflows are highlighted by the red crosses. (d) Histogram of the distribution of the measured  initial (dark gray) and final (light gray) speeds (in projection) of the plasma upflows, plasma downflows, and contracting loops. An animation is available for the \textit{SDO}/AIA 131 \AA\ background-detrended images and its time-distance plot. \label{fig:outflows}}
\end{figure*}

\subsection{Bi-directional outflows} \label{sec:outflows}
Shortly after the eruption of the dark cavity at around 15:54 UT, a thin bright plasma sheet appeared in multiple \textit{SDO}/AIA passbands, with a temperature of $\sim15$--20 MK according to EUV spectroscopic data \citep{2018ApJ...854..122W}. In SDO/AIA 131 \AA\ (which is sensitive to the \ion{Fe}{21} line at $\sim$10 MK; \citealt{2010A&A...521A..21O}) images, multitudes of plasma outflows are present in the plasma sheet during different phases of the event for an extended period of time \citep{2018ApJ...868..148L,2018ApJ...866...64C,2019ApJ...875...33H,Chen2020b,lee_formation_2020}. Here we focus on the plasma outflows during the post-impulsive gradual phase from 16:20 UT to 20:20 UT. We find many recurring pairs of bi-directional plasma outflows that propagate simultaneously in the sunward (down) and anti-sunward (up) direction. A time sequence of one such outflow is shown in Fig.\,\ref{fig:outflows}(a), in which the slow-varying background is removed to enhance the dynamic features. The upward-moving EUV outflows extend well into the \textit{MLSO}/K-cor field of view in white light to at least 1200 Mm (or 1.7 R$_\odot$) above the solar surface (Fig.\,\ref{fig:outflows}(b); \edit1{see also Figure \ref{fig:stackplot_appx} in Appendix \ref{sec:app} with a full height range}). Some of the upward-moving outflows are reported in a recent paper by \citet{lee_formation_2020} that reaching a height of more than 4 R$_\odot$ in \textit{SOHO}/LASCO C2 images. The downward-moving EUV outflows seem to terminate at the looptop region. Each pair of bi-directional outflows appears to diverge from a discrete site at varying heights in the plasma sheet. 

\begin{figure*}[ht!]
\epsscale{1.1}
\plotone{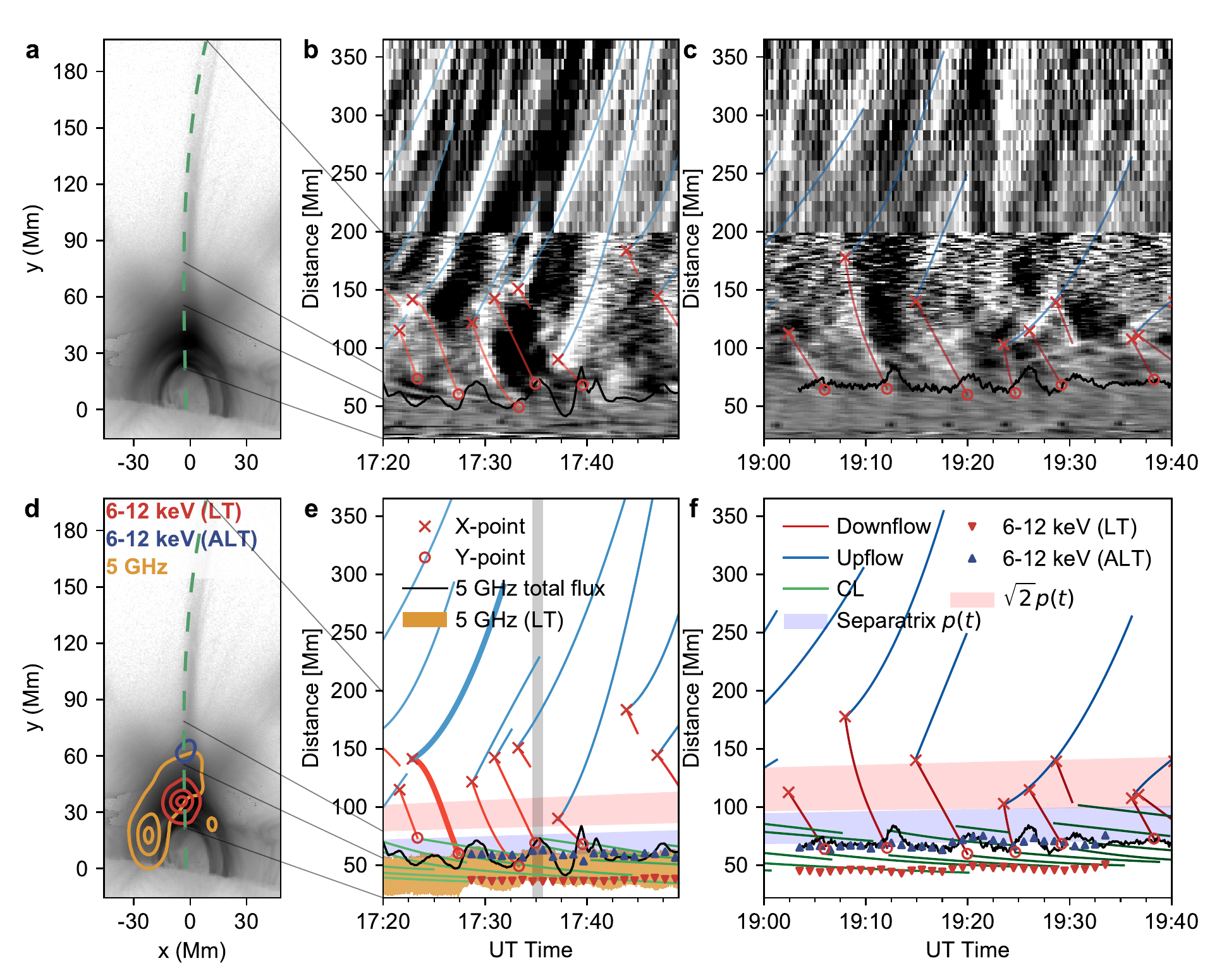}
\caption{(a) {\it SDO}/AIA 131 \AA\ image at 17:35 UT and the cut (same as the cut in Figure\,\ref{fig:flare}(b) and (c)). (b) and (c) Enlarged view of the time-distance plot for two selected time intervals shown as the white dashed boxes in Figure\,\ref{fig:outflows}(b). The tracks of outflows are denoted by blue and red curves. The possible X- and Y-point are denoted by cross and circle. The black solid line is the detrended \textit{EOVSA} 5 GHz light curve (same as the one shown in Figure\,\ref{fig:flare}(c)) superposed on the separatrix region that divides the fast downflows from the slow contracting loops. (d) Same as (a), but overlaid with 10\%, 50\%, and 90\% contours of the 5 GHz microwave source (yellow) and 6--12 keV X-ray looptop (LT) source (red) and the weaker ALT X-ray source (blue). (e) and (f) Tracks of outflows (blue and red) and contracting loops (CL; green) in the time-distance plots above. Also shown are the centroid locations of the looptop (red triangles) and ALT (blue triangles) X-ray sources, and the height range enclosed by the 10\% contour of the 5 GHz microwave source (orange). The timing of the 6-12 keV X-ray and 5 GHz microwave images shown in (d) are indicated by the gray-shaded vertical stripe in (e). \label{fig:stackplt_enlarge}}
\end{figure*}

To quantify the motion of the bi-directional plasma outflows, we construct a time--distance diagram from a slice along the direction of the plasma sheet feature seen in both {\it SDO}/AIA 131 \AA\ and {\it MLSO}/K-cor images (green dashed curve in Figs. \ref{fig:flare}(b) and (c)). The slice has a width $w_0 = 3\,\mathrm{Mm}$ at the base. To improve the signal to noise at larger coronal distances, we increase the width of the slice linearly with distance $d$ as $w(d) = w_0 + 0.04d$. At each time $t$, for every distance $d$ along the slice, all pixels across the slice within width $w$ are averaged to produce the intensity shown in the composite time--distance plot $I(t, d)$. The plasma upflows seen in EUV continue smoothly to the white light image seen at the upper edge of the \textit{SDO}/AIA field of view at $d\approx200$ Mm (Fig.\,\ref{fig:outflows}\,(b)). We selected the most prominent tracks and fitted either straight line or basis spline curve \citep{DEBOOR197250} to the projected height $h(t)$ as a function of time $t$, depending on their apparent curvatures in the time-distance map. We identified 40 pairs of such bi-directional outflow tracks in 16:20--20:20 UT. There are also a few additional cases of downflow tracks without an obvious upward counterpart. Fig.~\ref{fig:outflows}(d) shows the statistical distributions of the outflow speeds. The distributions for both the ``initial'' and ``final'' speeds, defined respectively as those measured when the flows appear and disappear (or indiscernible) in the time-distance maps, are displayed. The initial speeds of the upflows and downflows in projection are distributed between 100--900 $\rm{km\,s^{-1}}$ with an average of 250 km s$^{-1}$. These measured outflow speeds are consistent with previous reports of outflows in the same event \citep{2018ApJ...866...64C,2018ApJ...868..148L,2019ApJ...875...33H}, and are typical for SADs and SADLs reported in other events \citep{1999ApJ...519L..93M,2004ApJ...605L..77A,2009ApJ...697.1569M,2011ApJ...730...98S,2012ApJ...745L...6T, 2013ApJ...767..168L}.

The flow tracks in the time-distance map of Fig.~\ref{fig:outflows}(c) show that most downflows have nearly constant speeds along their path, while an upward acceleration is present in most upflows. The latter is also clearly shown in Fig.~\ref{fig:outflows}(d), in which the final speeds of the upflows are consistently greater than the initial speeds. Such an apparent acceleration of the upflows in the same event was also presented and discussed in previous studies \citep{2018ApJ...866...64C,lee_formation_2020}. The variation of the flow speeds along the plasma sheet may provide important hints for understanding the detailed physics within the reconnection current sheet including, possibly, the size of the diffusion region \citep{2018ApJ...858...70F}. However, a more in-depth investigation is beyond the scope of this study.

Similar to the interpretation adopted in previous studies \citep{2010ApJ...722..329S,2012ApJ...745L...6T,2013ApJ...767..168L}, we attribute the diverging location of each bi-directional outflow pair as the site of the reconnection ``X'' point associated with an individual magnetic reconnection event (or, to be more precise, the ``stagnation point'' of the reconnection outflows; see, e.g., \citealt{2018ApJ...858...70F}). Most of these identified reconnection sites are located at $d\approx50$--180 Mm (or 0.07--0.26 $R_{\odot}$) above the limb, which is only 1\%--3\% of the total length of the plasma sheet ($\sim$10 $R_{\odot}$) during that period. 
% \textbf{Note the $y$-axis of Fig.\,\ref{fig:outflows}(b,c) is displayed in {\it logarithmic} scale to accommodate the large height range in which the downflows and upflows are present within the same time-distance plot.} 

The downflows fade away as they merge into the tip of the cusp-shaped flare arcade (sometimes referred to as the ``Y'' point; \citealt{priest_forbes_2000,Chen2020b}), where numerous slow, downward-contracting loops are present (see the animation accompanying Fig.\,\ref{fig:outflows}). The slow contracting loops are also visible in the time-distance plots in Fig.\,\ref{fig:outflows}(b) as multiple faint, finer tracks that branch off from the faster downflow tracks. Their initial speeds, measured using the slopes of the green lines in the time-distance plots, are only $\sim$10s $\rm{km\,s^{-1}}$ or below. Such slow loop shrinkage is persistent throughout the gradual phase with an average recurrence period of $\sim$3.2 minutes. Although there is no one-to-one correspondence between the fast plasma downflows and the slow contracting motion of the post-reconnection flare loops, the slow-contracting loops appear in the close vicinity of the region where the fast downflows fade away, suggesting the presence of a ``separatrix'' region where the downflow motions appear to ``terminate''. %that they may be associated with the impulsive energy release therein. 

The location of this separatrix region between the downflows and the contracting loops nearly coincides with the tip of the cusp-shaped flare arcade, shown as the blue colored shading in Figure\,\ref{fig:outflows}(b--c), The lower edge of the separatrix region follows the end points of the fast downflow tracks. The upper edge follows the initial points of the slow contracting loop tracks. This separatrix region rises slowly during the gradual phase (blue shaded region in Figure\,\ref{fig:stackplt_enlarge}(e--f)) in a similar fashion as the slow rise motion of the underlying microwave and X-ray looptop source located at the top of the flare arcade (red triangles in Figure\,\ref{fig:stackplt_enlarge}(e--f); see also \citealt{2018ApJ...863...83G,2019ApJ...875...33H}). The centroids of the ALT 6--12 keV X-ray source (blue triangles in Figure\,\ref{fig:stackplt_enlarge}(e--f))   are located near the separatrix region and follow the same rising motion. Possible implications of such a spatial-temporal coincidence will be discussed in the next section.

To illustrate the timing of the impulsive microwave bursts in accordance with the observed EUV plasma downflows, we overlay the \textit{EOVSA} 5 GHz microwave light curve (from Figure\,\ref{fig:flare}(c)) on the time--distance plots in Figure\,\ref{fig:stackplt_enlarge} near the separatrix region. The arrival of most  plasma downflows at the separatrix region is immediately followed by a microwave burst. This correlation in both space and time is a strong indication for a casual connection between the plasma downflows arriving at the looptop and the appearance of microwave-emitting nonthermal electrons in the flare arcade.

\begin{figure*}[ht!]
\epsscale{1.2}
\plotone{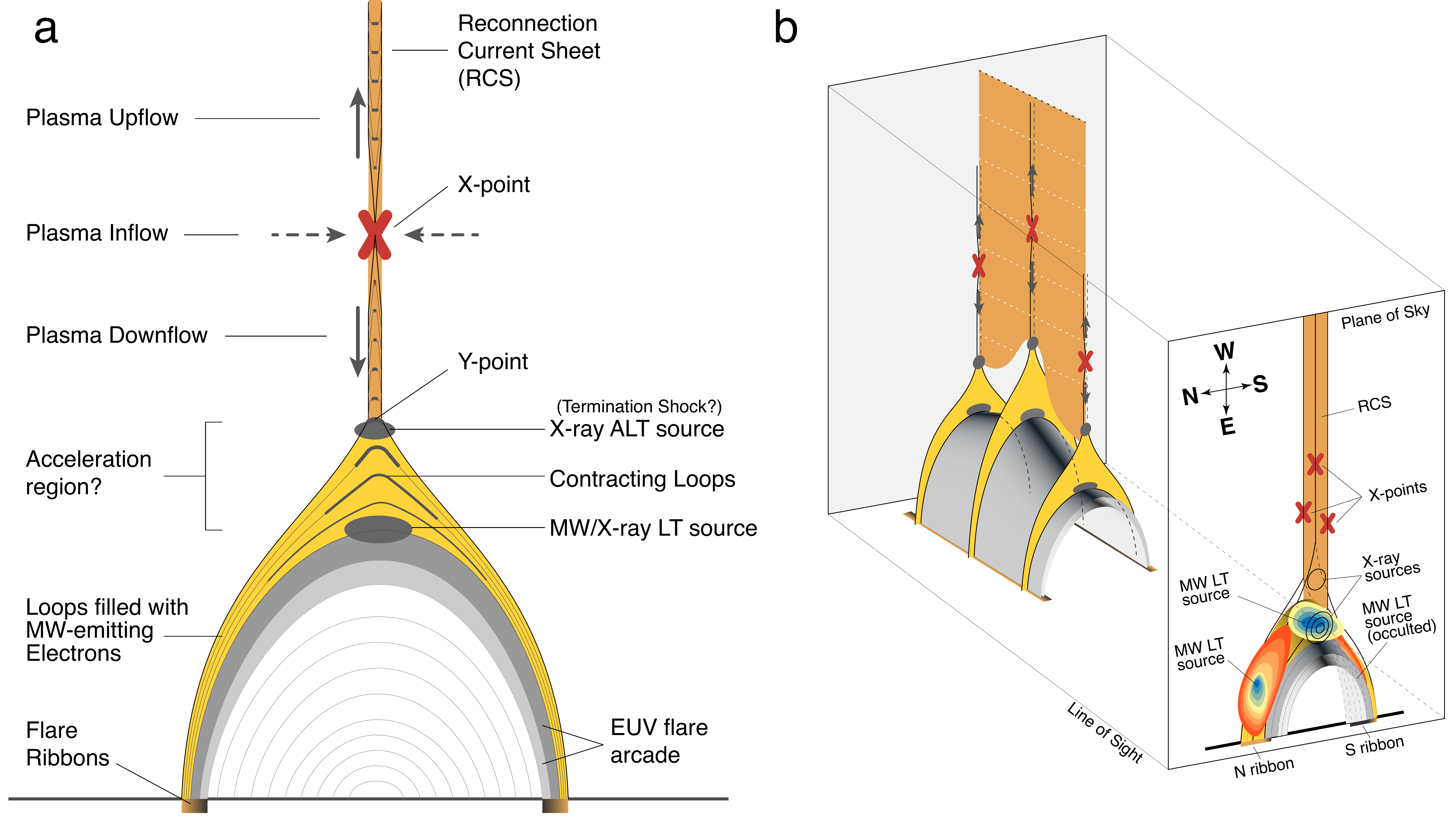}
\caption{(a) Schematic diagram of post-impulsive flare arcade and the large-scale reconnection current sheet with an edge-on view (adapted from \citet{1996ApJ...459..330F}).  Reconnection at multiple X points within the RCS results in a pair of highly- bent flux tubes that shrink quickly in both the sunward and anti-sunward directions, observed as the EUV plasma outflows. A microwave and X-ray source appears at the looptop due to plasma heating. Accelerated electrons in the flare arcade give rise to the nonthermal loopleg microwave source. A weak X-ray source is present near the Y point at the bottom of the RCS, where the fast downflows turn into slow contracting loops. The flare arcade itself is visible as a bright EUV arcade consisting of many strands. %Energy released in the X-point are carried by or moves along with the downflow toward the flare loop, resulting in a HXR burst due to plasma heating in the loop top region, and a microwave burst produced by nonthermal electrons accelerated within or ejected into the cusp loop.
(b) Schematic diagram of flare arcade and the RCS depicted in 3D. Discrete reconnection events occur at different times and heights within the 3D RCS, visible as the observed scattering of the reconnection sites viewed edge on. Schematic of the observational signatures including the plasma sheet (with a finite width), EUV flare arcade, as well as microwave and X-ray sources are shown projected on the plane of sky. Here we adopt the possible interpretation in which the flare arcade may be slightly tilted with respect to the line of sight, which may account for the absence of the microwave source in the southern (right) side of the arcade.  \label{fig:cartoon}}
\end{figure*}

\subsection{Summary of the Observations} 

The main observational findings discussed in this section are briefly summarized as below: 

\begin{enumerate}
\item Reconnection sites (or ``X'' points), from where the bi-directional plasma outflows diverge, reside at low altitudes $<$180 Mm, which are  $\sim$1--3\% of the total length of the long plasma sheet, which extends to $>$10$R_{\odot}$.

\item The arrival of most EUV plasma downflows at the top of the cusp-shaped flare arcade correlates with an impulsive microwave and X-ray burst, which consists of a (mostly) thermal looptop microwave and X-ray source and a nonthermal loopleg microwave source. The individual loopleg microwave nonthermal bursts and the looptop X-ray bursts take place simultaneously within a uncertainty of $\sim$10--20 seconds.

\item Multitudes of slow contracting loops are present below the tip of the cusp-shaped flare arcade where the fast plasma downflows terminate. A secondary ALT X-ray source coincides with the separatrix region that divides the fast downflows and slow contracting loops.

\end{enumerate}

\section{Discussion and Conclusion} 
\label{sec:dis}

Our observational results are consistent with the standard CSHKP eruptive flare scenario for the post-impulsive phase (or gradual phase). At this stage, the eruption has already propagated to a remote coronal distance, leaving behind a large-scale vertical RCS above the post-flare arcade (see, e.g., \citealt{2018ApJ...858...70F}). %We propose the following physical picture that ties together the observational results. 
Figure\,\ref{fig:cartoon}(a) shows a schematic diagram for the post-impulsive phase projected in 2D adapted from a well-known cartoon in \citealt{1996ApJ...459..330F}. In this cartoon, the RCS and the underlying flare arcade are viewed edge on, in accordance with the viewing perspective of this event. Sporadic magnetic reconnections occur at localized magnetic null points (or X points) in the RCS, creating pairs of highly bent magnetic flux tubes \citep{1963PhFl....6..459F}. Plasma is ejected from the X points both upward and downward along the RCS, resulting in bi-directional plasma outflows. %at They drag plasmoids at their apex away from the X-point under their magnetic tension force at a fraction of the Alfv\'en speed \citep{2018ApJ...868..148L}. The plasmoids have a density enhancement against the background, observed in the forms of bright plasma upflow and downflow along the CS. The bent flux tube above the X-point moves unhindered towards the CME. The other set of flux tube below the X-point moves towards the cusp-shaped flare arcade in a similar fashion until it arrives at the tip of the flare cusp, or the Y-point. The flux tube relaxes to less bent shape. The plasmoid trapped at the apex of the bent flux tube diffuses along the field lines and moves with it downwards, observed as a contracting loop, which undergoes strong deceleration as it impinges the flaring loops piled up underlying in the cusp region (lines in the yellow region in Fig.\,\ref{fig:cartoon}a). 
% The average speed of the contracting loops drops to 10s km $\rm{s^{-1}}$ from 100s km $\rm{s^{-1}}$ as of the downflows. 

The reconnection sites, pinpointed by the bi-directional plasma outflows, are located very low in the RCS. The heights of the X points are  $\sim$1--3\% of the total length of the RCS seen in EUV and white light, which extends to at least 10 $R_{\odot}$. The observed low reconnection sites within a long RCS are in  agreement with  predictions in the 2D theoretical model by \citet{2018ApJ...858...70F}, the latest development based upon one of most well-known standard flare models in \citet{2000JGR...105.2375L}, \citet{2005ApJ...630.1133R} and \citet{2009ApJ...701..348S}. The 2D theoretical model predicts the height of the X point of approximately $\sqrt{2}p(t)$ during the post-impulsive phase, where $p(t)$ is the height of the Y point at the lower end of the RCS at time $t$. The latter marks the location where a thin RCS turns into a cusp-shaped post-reconnection flare arcade, measured as the tip of cusp loops seen in \textit{SDO}/AIA 131 \AA\ time-series images. As discussed in the previous section, this location also coincides with the separatrix region where the fast downflows, identified in the thin plasma sheet, meet the slow-contracting cusp loops below the cusp tip. In Figs.\,\ref{fig:outflows} and \ref{fig:stackplt_enlarge}, we show the estimated location of the rising Y point as the blue shaded region and, according to the prediction in \citet{2018ApJ...858...70F}, the presumed location of the reconnection X point $\sqrt{2}p(t)$ as the red shaded region. The predicted X point in idealized 2D flare model and its evolution in time agree with the location of the reconnection events pinpointed by the bi-directional plasma outflows in the thin plasma sheet. The scatter of the observationally-inferred reconnection sites around the model-predicted X point location is likely due to a deviation of the actual reconnection events from the idealized 2D model, together with the 3D nature of the flare event with multiple reconnection events distributed within the extended RCS along the direction of the LOS (illustrated in Fig.\,\ref{fig:cartoon}(b)).  

%show the predicted height altitude of the lower tip of a long CS in the post-impulsive of a solar flare. The low altitude of X-point can easily account for the quasi-steady reconnection process in the stable CS structure, as the reconnection rate are largely controlled by geometry of the magnetic field just above the flare arcade, which change very slowly in time during the decay phase \citep{2018ApJ...858...70F}. The compact size of X-points and their low altitude in the current, also suggests that reconnection does not resemble the Sweet--Parker model (ref) that invokes a long diffusion region throughout the CS, but resembles to Petschek-like reconnection. 

In the 2D reconnection theories, the speeds of the reconnection outflows are at the local Alfv\'en speed \citep{1958IAUS....6..123S,1957JGR....62..509P,1967ApJ...147.1157P}. However, similar to many other reports \citep{2010ApJ...722..329S,2011ApJ...730...98S,2018ApJ...868..148L,2019ApJ...875...33H,Chen2020b}, in our observations, the speeds of the bi-directional plasma outflows are between 100--900 km s$^{-1}$ with an average of 250 km s$^{-1}$, which are likely sub-Alfv\'enic. It has been suggested that the observed plasma outflows may be sub-Alfv\'enic due to 3D effects, or have been slowed down as they emerge from the reconnection diffusion region by, e.g., an aerodynamic drag force \citep{2018ApJ...868..148L}. 

The plasma outflows carry a significant portion of the total released magnetic energy in the form of electromagnetic Poynting flux, enthalpy flux, and kinetic energy flux of the bulk flows and turbulence \citep{2008ApJ...675.1645F,2009ApJ...695.1151B,2010ApJ...721.1547R,2017PhRvL.118o5101K,2018ApJ...854..122W,2018ApJ...864...63P,2018ApJ...866...64C}. Arrival of the downward-propagating plasma outflows at the cusp region dissipates their energy, resulting in plasma heating through thermal conduction and/or adiabatic heating \citep[see, e.g., recent 3D modeling results in][]{2019ApJ...887..103R}. If a fast-mode termination shock is established in the cusp region (which is perhaps implicated by the presence of the secondary ALT X-ray source near the cusp tip), plasma heating would occur in the shock downstream region \citep{1986ApJ...305..553F,1994Natur.371..495M}. Such heated plasma is revealed by the thermal X-ray and microwave source observed at the looptop. 

The impulsively released magnetic energy during the sporadic magnetic reconnection events in the RCS can also lead to particle acceleration. Electrons can be accelerated to nonthermal energies in the RCS, at the looptop, or in the flare arcade itself by a variety of acceleration mechanisms (see, e.g., \citealt{1997JGR...10214631M} for a review). In our observations, the nonthermal microwave bursts only occur in the loopleg region at a large distance away from the reconnection X points. 
This is in line with HXR and microwave imaging spectroscopy data \citep{2010ApJ...714.1108K,2014ApJ...780..107K,Fleishman2020,Chen2020b} providing increasing evidence favoring the looptops/cusp regions as the primary electron acceleration sites.  

In this event, the cusp region as the primary electron acceleration site during the time of interest of our study is supported by the relative timing between the X-ray/microwave bursts and the magnetic reconnection events in the RCS (inferred from the occurrence of the bi-directional outflows). An important clue, as shown in Fig.\,\ref{fig:stackplt_enlarge}, is that the occurrence of the X-ray/microwave bursts correlates with the \textit{arrival} time of the plasma downflows at the cusp, but not the time of the magnetic reconnection events themselves. A straightforward interpretation is that the electrons responsible for the nonthermal microwave bursts are accelerated locally at the looptops, where freshly injected energy is available from the arrival of the plasma downflows. An alternative scenario is that the microwave-emitting electrons are accelerated in the RCS, but are trapped in the propagating plasma downflows. These electrons are released once the plasma downflows have arrived at the looptop, resulting in the nonthermal microwave sources. Although our data alone can not distinguish between the two scenarios, the latter scenario requires an additional mechanism that traps and/or accelerates the electrons within the plasma outflows along their path, while ``breaks'' this trapping upon the arrival of the outflows at the looptop. The cusp region is also favored as a site for direct plasma heating, implied by the concurrent appearance of the thermal emission seen in X-rays ($\lesssim$25 keV) by $\sim$20 MK plasma at the looptop, together with the nonthermal microwave emission.

% [\textbf{Option 1}:{\color{brown}We note that these reconnection-generated downflows may also be the driver of the persistent QPPs reported in \citet{2019ApJ...875...33H} with a period of $\sim150$ seconds, which was interpreted as MHD ocsillations in the flare arcade.}]

% [\textbf{Option 2:}{\color{brown} 
The observed recurring bursts possess a recurrence period of $\sim300$ seconds and a modulation depth of 1.5\% to 30\% reminiscent of quasi-periodic pulsations (QPPs) observed in X-rays and microwaves during flares \citep{2005A&A...440L..59F,2006A&A...460..865M,2011A&A...525A.112R}. Such QPPs with long periods are sometimes referred to as \textit{long} QPPs \citep{2009SSRv..149..119N}. One of the most intriguing questions about QPPs is what drives them during flares. The possible causes are generally categorized into two groups: (1) time-dependent energy release \citep{Fleishman2008a,2019ApJ...886L..25Y}, and 2) MHD oscillations in flare sites \citep[see][for reviews]{2009SSRv..149..119N,2018SSRv..214...45M}. Differentiating observationally between the possible explanations of long-period QPPs during flares remains elusive. The relative timing between X-ray/microwave bursts and the reconnection events we observe, however, allows for the determination of the origin of these particular bursts, although we leave open the question of whether they should be considered as QPPs. We find that the timing of the microwave and X-ray bursts is entirely driven by the reconnection at the X points, which rules out the MHD interpretation as the driver of the bursts. Nevertheless, \citet{2019ApJ...875...33H} recently reported persistent QPPs with a two-times shorter period of $\sim$ 150 seconds in \textit{GOES} SXR and \textit{SDO}/AIA 131 \AA\ light curves during the gradual phase of this same flare. The time-dependent magnetic reconnection mechanism we describe cannot fully account for these shorter-period QPPs, and other mechanisms such as MHD oscillations could play a role in modulating these lower-energy thermal emissions. As pointed out by \citet{2019ApJ...875...33H}, persistent QPPs require a renewed excitation of MHD modes in the flare arcade, and the recurring reconnection-generated downflows we describe could provide the needed repetitive trigger at or above the the post-flare arcade  \citep{2016ApJ...823..150T,2017ApJ...847...98J}.

In the ideal 2D standard model, the microwave source is expected to \textit{bestride} the flare arcade. However, there is a marked asymmetry in the observed microwave emission in the post-flare arcade: a microwave source is only visible in the northern loopleg while the southern leg is not. One explanation could be due to the perspective effect that the event is tilted slightly away from the ideal 2D standard model projection, so that the microwave source at the southern leg of the flare arcade may be occulted by optically-thick, dense plasma filled in the arcade (perhaps due to chromospheric evaporation) located in the forefront along the LOS (illustrated in Fig.$\,$\ref{fig:cartoon}(b)). However, we could not completely rule out a coronal loop asymmetry as a possible cause of the dissimilar radio brightness of the northern and southern loop legs. %Whether this is primarily due to suppression of microwave emission or a relative lack of nonthermal electrons in the northern leg is beyond the scope of the current study and remains to be determined by more in-depth spectral analysis now underway.

% To conclude, the performed multi-wavelength analysis allowed us to quantify particle acceleration and plasma heating in the post-flare arcades, observed as the hot and nonthermal flare emissions, and their relationships the magnetic reconnection in the RCS.
To summarize, thanks to the new microwave imaging spectroscopy observations from \textit{EOVSA}, we have presented a comprehensive study that associates the bi-directional EUV plasma outflows in a large-scale RCS to looptop microwave and X-ray bursts in both time and space. The performed multi-wavelength analysis allowed us to quantify particle acceleration and plasma heating in the post-flare arcades, observed as the hot and nonthermal flare emissions, and their relationships with the magnetic reconnection in the RCS. Our findings reveal new facets of magnetic reconnection, the subsequent energy conversion, and electron acceleration, and thus help to better understand these fundamental phenomena.

\acknowledgments
{EOVSA operation is supported by NSF grant AST-1910354. S.Y., B.C. are supported by NSF grants AGS-1654382, AGS-1723436, and AST-1735405 to NJIT. K.R. is supported by NSF grant AGS-1923365 to SAO and grant 80NSSC18K0732 from NASA to SAO. G.F. and G.N. are supported by NSF grant AGS-1817277 %New Flare NSF Grant
and NASA grants
80NSSC18K0667, %(H-GIO Cold Flare grant)
80NSSC19K0068, %(Dale's LWS grant)
80NSSC18K1128, %(Dale's H-SR grant)
to New Jersey Institute of Technology. The work was supported partly by NASA DRIVE Science Center grant 80NSSC20K0627. We thank Dr. Chengcai Shen for his inspiring MHD modeling on solar flares. We thank the anonymous referee who provided constructive comments to improve the paper. We thank the SDO/AIA team for providing the EUV data. We thank the \textit{RHESSI} and \textit{Fermi}-GBM team for providing the hard X-ray data. We also thank the \textit{SoHO}/LASCO and \textit{K-Cor} teams for providing white-light data.}

\facilities{OVRO:SA, SDO, SoHO, K-Cor, RHESSI, Fermi}

\vspace{2in}

\appendix
\label{ap}
\section{Plasma Upflows Extending to Very High Altitudes}
\begin{figure*}[!ht]
\epsscale{1.2}
\plotone{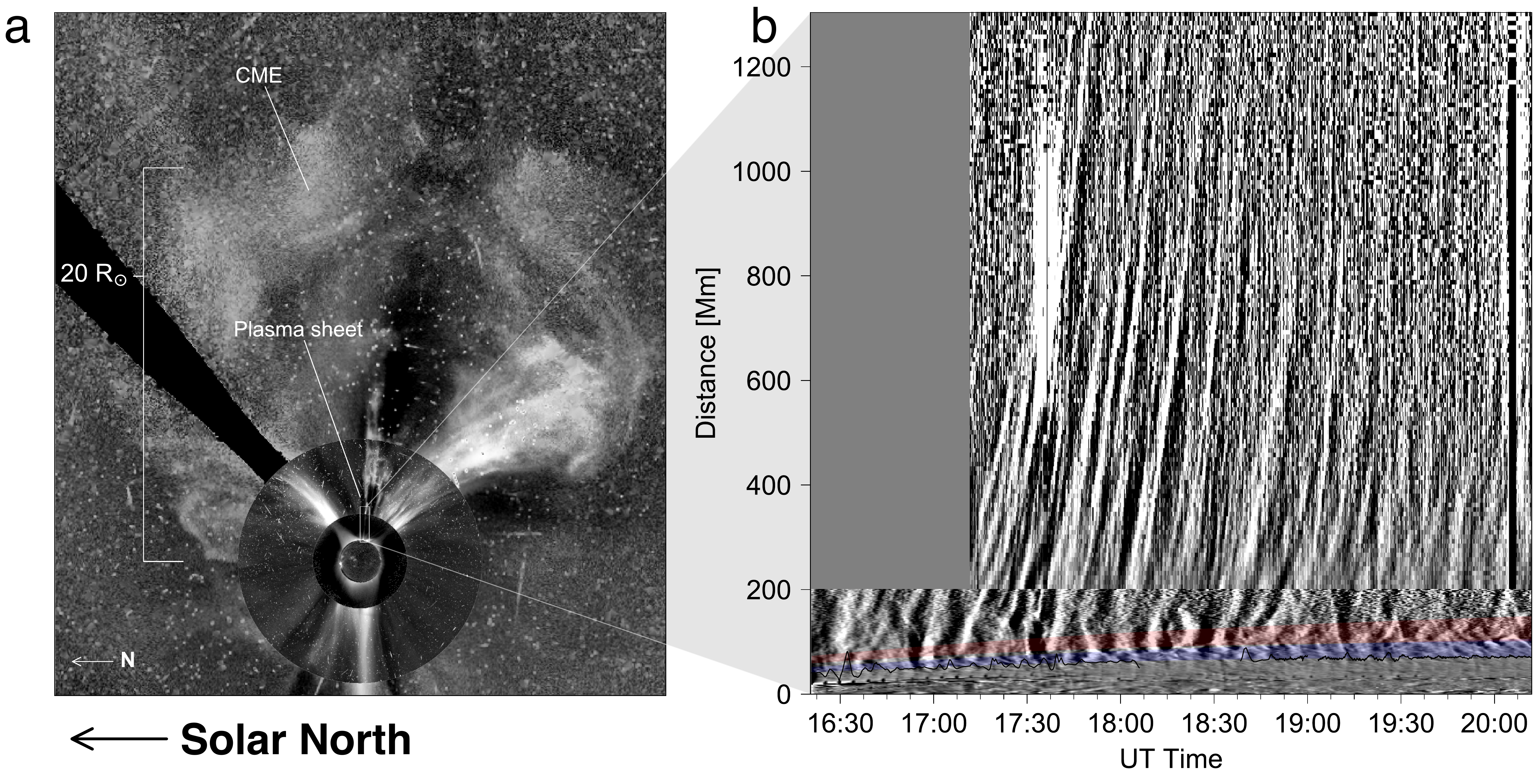}
\caption{Upflows are seen to extend to at least 1200 Mm (or $\sim$1.7$R_{\odot}$) above the solar surface in white light images. (a) Same as Fig.\,\ref{fig:flare}(a) (but in grayscale). (b) Same as Fig.\,\ref{fig:outflows}(b), but showing upflows that extend to much greater heights. \label{fig:stackplot_appx}}
\end{figure*}
\label{sec:app}
\edit1{The reconnection X points inferred from the diverging bi-directional EUV plasma outflows reside at a low altitude of $<$180 Mm. Multitudes of plasma downflows are present below the X points. Meanwhile, the EUV plasma upflows extend well into the \textit{MLSO}/K-cor's field of view in white light to at least 1200 Mm. In Figure\,\ref{fig:outflows}(b), we have limited the $y$-axis to $<$400 Mm for better showing the details of the downflows together with the upflows. For completeness, in Figure \ref{fig:stackplot_appx}, here we include a similar plot showing the full extend of the upflows up to $>$1200 Mm.}

\vspace{4in}

\bibliography{reference}{}
\bibliographystyle{aasjournal}

% \allauthors

\listofchanges

\end{document}